\documentclass[11pt,twoside]{article}

\usepackage{here}
\usepackage{graphicx}
\usepackage[footnotesize,nooneline,center]{caption2}
\usepackage{epsfig}
\usepackage{color}
\usepackage[vflt]{floatflt}
\usepackage{amssymb,amsmath}
\usepackage{rotating}
\usepackage{fancyhdr}

\definecolor{d}{rgb}{0,0.7,0}
\definecolor{s}{rgb}{0,0,0.7}

\def\abs#1{|#1|}

\def \CC {{\mathbb{C}}}
\def \RR {{\mathbb{R}}}

\def \QQ {{\mathbb{Q}}}

\def \NN {{\mathbb{N}}}

\def \DD {{\mathbb{D}}}

\def \riemannsphere {\hat{\CC}}

\newcounter{theo} 

\newtheorem{definition}[theo]{\small\textbf{Definition}\normalsize}

\newtheorem{question}[theo]{\small\textbf{Question}\normalsize}

\numberwithin{equation}{section} \numberwithin{theo}{section}

\begin{document}
\title{One approach to the digital visualization\\of hedgehogs in holomorphic dynamics}
\author{Alessandro Rosa}

\maketitle

\begin{abstract}
In the field of holomorphic dynamics in one complex variable,
hedgehog is the local invariant set arising about a Cremer point
and endowed with a very complicate shape as well as relating to
very weak numerical conditions. We give a solution to the open
problem of its digital visualization, featuring either a time
saving approach and a far-reaching insight.
\end{abstract}

\section{Introduction}\label{Intro}
Let $\CC$ be the complex plane. Since $\infty$ cannot be handled
like an ordinary point here, as deserved by many complex
functions, this lack is fulfilled by the compactification
$\CC_\infty = \CC\cup\{\infty\}$, namely the \emph{extended
complex plane}; anyway the representation of the neighborhood of
$\infty$ is still impracticable here. Riemann cracked the problem
by a stereographic projection of $\CC_\infty$ onto a sphere of
radius 1: this is the Riemann sphere
$\riemannsphere$.\vspace{0.4cm}

Holomorphic dynamics collect the studies on the iterates $f_n$ of
the function $f:\CC_\infty^v\rightarrow\CC_\infty^v$ of given type
(entire, meromorphic, transcendental, \dots) and in one or several
complex variables (depending on $v\geq 1$). The map $f$ is of
finite degree.

We will deal with the case of one variable, where $v=1$. The
questions in this field may rise to high degrees of complication
and many ones deserve a multilateral attack rooting into Complex
Analysis, Topology, Theory of Numbers, Uniformization Theory. One
of the most important goals is the study of elements not changing
under iteration: the so-called \emph{invariants}, showing up to
dimension $2$ at most: they might be points (dim 0), lines (dim 1)
or surfaces (dim 2). The collection of invariants with a same
property is the \emph{invariant set}.

It is convenient to split results into the `\emph{local}' and into
the `\emph{global}' branch, in order to have a good picture of the
whole corpus. A local study investigates on the properties which
hold up to a finite distance from the invariant set, thus inside a
bounded domain; while a global approach wants those properties
enjoyed by points being even at infinite distance, thus all over
the $\riemannsphere$. Branches are not disjoint and related
concepts act in mutual cooperation: in fact, either locally or
globally, the fate of (inverse and forward\footnote{Whether the
iterative index $n$ is negative or positive respectively.}) orbits
closely relates to the nature of the fixed point $\delta$ and its
neighborhood.

Results rely on the study of the \emph{orbits}, i.e. the set of
points generated by iterating the given function $f$,
$$f_1\equiv f, f_2\equiv f[f], \dots, \ f_n\equiv f[f_{n-1}].$$
approaching to limit sets which may consist of finitely (dim 0),
or of infinitely and uncountably many (dim 1 and 2) points. For
such sets finitely many points of order $k>1$, one speaks of
\emph{limit cycle of periodic points} $\delta_n$:
\begin{equation}\label{periodiccycle}
f(\delta_1)=\delta_2, f(\delta_2)=\delta_3, \dots,
f(\delta_{k-1})=\delta_1.
\end{equation}
If the period is $k\geq 1$, then $\delta_{1\leq i\leq
k}=f_k(\delta_{1\leq i\leq k})$. If $k=1$, we have the \emph{limit
fixed point} $\delta$ and the expression \eqref{periodiccycle}
boils down to:
\begin{equation}\label{fixedpoint}
\delta=f(\delta).
\end{equation}
\noindent so the fixed point can be re-framed as a cycle of period
$1$.

Other cycles may belong to invariants $J$ of dimension 0
(infinitely many points), or 1 and, exceptionally, of
dimension\footnote{In this latter case, they span all over the
Riemann sphere: $J\equiv\CC_\infty$.} 2 (both, uncountably many).

In the economy of dynamics over $\riemannsphere$, cycles of
finitely or of infinitely and of uncountably many points play
different roles which are explained locally in the former case,
while the latter cycles are object of the global investigation.

In the second case, we speak of `\emph{Julia sets}' $J$: when they
include infinitely many points, their topology is totally
disconnected; when uncountably many, they are continuous (Jordan
curves or not).

With the caution to the casuistry of related local dynamics, a
same limit cycle of finitely many points groups all orbits
converging to it into macro-sets, said `\emph{the basins}
$\mathcal{B}$ \emph{of attraction}', also defined `\emph{Fatou
sets}', in honor of Pierre Fatou (1878--1929), who co-pioneered
these investigations in the same times (1918--20) and
independently from Gaston Julia (1893--1978), who is credited as
the first official discoverer\footnote{(Related historical events
are not fair.) Each basin relates to a same limit cycle. A fuller
historiographic investigation will appear in \cite{AIR}.} of sets
$J$ in 1918. While the role of the basins is to include all those
orbits to a same limit cycle, $J$ is the boundary between the
basins.

Julia sets are well-known objects at all levels today, sparkling
the imagination a wide range of people, from artists to
mathematicians. The details of their suggestive and very
complicated fractal shapes were disclosed to human eyes by the
early computer experiments during late 1970s: machines revealed to
be the indispensable aid for overcoming the long standing barriers
of hand-written, rough plots available to those ancient
mathematicians. The technologic run is continuously opening to
finer results, thanks to higher screen resolutions and faster
{\small CPU} saving from long time consuming computations, as
required to render these images. Anyway this side role shows that
Technology is not a priority and does not fully belong to the
road-map of this field: it is just a good tool to develop those
methods mastering both analytical and geometrical relations --
which are most wanted today --, together with more accurate
numerical precision, especially for the question we are going to
deal with.

Even at the graphical level, we differ local from global methods,
whether they focus on the previous limit sets related to the two
branches.

After some introductory theory culminating into the presentation
of what the invariant sets $\mathcal{H}$, said `\emph{hedgehogs}',
are meant in holomorphic dynamics, we will illustrate the related
problem by showing how most available techniques fail to display
$\mathcal{H}$ adequately; finally, we will discuss how to solve
this question and to code an approach via
pseudo-C$^{\textnormal{++}}$ language code.

\section{Basic theory}\label{IntroductoryTheory}
\subsection{The four cases}\label{FourCases}
We are going to sketch out the mathematical terms of our local
environment: here complex dynamics are mostly interested in the
orbits behavior induced by iterates $f_n(z)$ near limit cycles of
the map $f$ and focuses on cycles of order 1, i.e. fixed points
$\delta, f(\delta)=\delta$. For an easier approach, we will assume
fixed points from now on. Iterates are operators defined as
follows in the forward sense ($n\geq 0$):
$$f_0(z)\equiv z, f_1(z)\equiv f(z),f_2(z)\equiv f[f(z)],\dots\ ,f_n(z)\equiv f[f_{n-1}(z)];$$
\noindent or backwards by a composition of inverse maps ($n<0$):
$$f_{-1}(z)\equiv f^{-1}(z), f_{-2}(z)\equiv f^{-1}[f^{-1}(z)],\dots\ .$$

\noindent The starting point $z$ of an orbit is said the \emph{seed}.

The classification of $\delta$ is achieved by computing
the modulus $\lambda$ of the first derivative $f'$ at $\delta$,
$\lambda=\abs{f'(\delta)}$ and it is essential to understand the
nature of local invariant sets, grouped here into four main
classes:

\begin{enumerate}\label{FixedList}
    \item \emph{super-attracting} fixed point, when $|\lambda|=0$;\label{FixedPointSuper}
    \item \emph{attracting} fixed point, when $0<|\lambda|<1$;\label{FixedPointAttracting}
    \item \emph{indifferent} or \emph{neutral} fixed point, when $|\lambda|=1=|e^{2\pi i\theta}|$;\label{FixedPointNeutral}
    \item \emph{repelling} fixed point, when $|\lambda|>0$;\label{FixedPointRepelling}
\end{enumerate}

Reader's understanding can be lessened by splitting this basic
classification into two extremal cases, (super--)attracting and
repelling entries, their directions are opposite but dynamical
properties are very similar. With it, one understand that the
indifferent case stands at the middle way and it is the
conjunction point among those ones, either in a \emph{inclusive}
way by presenting subcases where all dynamical features are shown
at the same time and in an \emph{exclusive} way, i.e. showing
features which are not enjoyed by both extremal cases.

In fact (super--)attracting fixed points can be reached by forward
iterates, whereas Julia sets include cycles of repelling points
and reached by backward iterates. Obviously, the dynamical
characters (contraction, repulsion) of entries
\ref{FixedPointSuper}), \ref{FixedPointAttracting}) and
\ref{FixedPointRepelling}) hold for the direct function $f(z)$,
while they are inverted for $f^{-1}(z)$. Indifferent points
deserve a separate discussion, where even the concept of `limit'
shall be carefully applied, because not conventionally intended
like in the other cases.

While entries \ref{FixedPointSuper}), \ref{FixedPointAttracting})
and \ref{FixedPointRepelling}) just need the modulus value, case
\ref{FixedPointNeutral}) requires a thorough investigation: the
modulus is insufficient to distinguish the more complex casuistry
here, as illustrated in table \ref{IndifferentTable}; another
parameter is demanded and our next (and immediate) choice is the
angle $0\leq\theta<2\pi, \theta\in\RR$. Its numerical properties
rule out the dynamical characters of neighboring orbits about
indifferent points. The main separation, into the two great
classes of $\theta\in\QQ$ (rationally indifferent $\delta$,
\emph{parabolic} case) and of $\theta\in\RR\backslash\QQ$
(irrationally indifferent $\delta$, \emph{elliptic} case), is
followed by a number of sub-cases. The former relates to one only
invariant set, namely the Fatou-Leau flower (see def. at p.
\pageref{DefFatouLeauFlower}), and does not branch out; while
$\theta\in\RR\backslash\QQ$ generates a richer variety whose local
dynamics get far more complicate: here a second level opens to
local invariant sets, distinguishing for the numerical properties
of $\theta$, some of which are extremely weak and shunning the
machine finite digits computation. The goal of this work is
understand how and if such latter cases might be actually
attackable in particular graphical terms which evince their local
dynamics.

\subsection{Analogy with linear models}\label{AnalogyModels}
One crucial tool to study the local dynamics is the
\emph{Schr\"{o}der functional equation} (SFE):
\begin{equation}\label{SFE_general}
\psi[f(z)]=a[\psi(z)],
\end{equation}
where $\psi(z)$ is an invertible map. Without loss of
generalization, let the origin be fixed for $f$. If we replace
$a(z)$ with $\lambda\equiv f'(z)$ -- the meaning of the first
derivative strengthens the application of SFE to local problems,
then \eqref{SFE_general} turns into this version:
\begin{equation}\label{SFE_holodyns}
\psi[f(z)]=\lambda\psi(z).
\end{equation}

Given a sufficiently close neighborhood $D$ of $\delta$, the left
diagram below, as well as the right one extending to iterates,
commutes when SFE holds:
\begin{table}[h]
  \centering
    \begin{tabular}{cccp{2.0cm}ccc}
    $D$ & $\xrightarrow{}$ & $f(D)$ & & $D$ & $\xrightarrow{}$ & $f^n(D)$\\
    $\psi\downarrow$ & &$\psi\downarrow$ & & $\psi\downarrow$ & &$\psi\downarrow$\\
    $\CC_\infty$ & $\xrightarrow{\lambda}$ & $\CC_\infty$ & & $\CC_\infty$ & $\xrightarrow{\lambda^n}$ & $\CC_\infty$\\
    \end{tabular}\label{SFEdiagram}
\end{table}

The major goal of \eqref{SFE_holodyns} is to set up an analogy
between the behavior of iterates $f_n(z)$ about $\delta$ and an
easier model; so, if SFE commutes, there is a local change of
coordinates where $f(0)=0, f'(z)=\lambda$: the local behavior of
orbits $f_n(z)$ can be studied by means of the simpler
investigation of the linear map $\lambda^n z$.

\begin{table}[h]
  \centering
  \input{hedge_shot_34.pic}
  \setcaptionwidth{0.9\textwidth} \captionstyle{normal}\caption{\textbf{Resuming the neutrality}. A diagram
  illustrating the classification of all dynamics for indifferent fixed
points.}\label{IndifferentTable}
\end{table}

Local theory had proven that SFE holds only for the
(super--)attracting and the repelling case, together with the
rationally indifferent and part of the irrationally indifferent
cases (see list at p. \pageref{FixedList}). If the local change of
coordinates into $\lambda z$ applies, $D$ is a
\textbf{B\"{o}ttcher} or a \textbf{Koenigs' domain}, depending on
$\lambda=0$ and $\lambda<1$ respectively. For $\lambda>1$, local
dynamics are repelling and, locally speaking exclusively, this
case can be regarded as the converse of the attracting one; anyway
here Fatou and Julia showed that it makes more
sense\footnote{According to their researches, one evinces the
boundary role of repelling fixed points and cycles in the global
dynamics all over the Riemann sphere. The former researchers did
not score this goal, because they regarded repulsion like the
converse of attraction.} of studying repelling orbits and the
related invariant set (as resumed in diagram \ref{MotionTable}) in
global terms: this is the \textbf{Julia set}, defined as \emph{the
closure of all repelling cycles} and being the common frontier of
all basins $\mathcal{B}$ of convergence.

\begin{table}[h]
  \centering
  \input{motion_diag.pic}
  \setcaptionwidth{0.8\textwidth} \captionstyle{normal}\caption{\textbf{Classification by motions}.
  This is the classification of invariant sets for holomorphic dynamics in one complex variable by
  the three elementary motions in the plane.}\label{MotionTable}
\end{table}
The discussion of $|\lambda|=1$ requires a much more delicate
analysis: first, its re-writing in the formal terms of
\eqref{SFE_holodyns} is
\begin{equation}\label{SFE_Siegel}
\psi[f(z)]=e^{2\pi
i\theta}\psi(z)\quad\Rightarrow\quad\psi[f_n(z)]=e^{2\pi i\theta
n}\psi(z),
\end{equation}
where the version on the right refers to iterates $f_n(z)$,
conjugated locally to the family of complex rotations
$e^{2\pi\theta in}z$. Problems arose when it was seen that
\eqref{SFE_Siegel} does not always commute, given $\theta\in\RR$,
whether it does or not is up to the numerical nature of $\theta$.
For example, it never commutes for $\theta\in\QQ$, while it does
when $\theta\in\RR\backslash\QQ$ satisfies this Diophantine
condition: given two numbers $r>0,k\geq 2$,
\begin{equation}\label{DiophantineCond}
\abs{\theta-p/q}>r/q^k,
\end{equation}
for every rational number $p/q$ where $p,q\in\NN$. Then $f_n(z)$
are said to be \emph{linearizable} into a complex rigid rotation
or, simply, `linearizable'; topologically speaking, there exists a
neighborhood of $\delta$ which is conformally isomorphic to a disc
and where the dynamics of $f_n(z)$ are rotational, the
\textbf{Siegel disc}. When \eqref{DiophantineCond} holds, $\theta$
is \emph{poorly approximated} by rational numbers; otherwise it is
\emph{goodly approximated}. It is clear that rational numbers
(parabolic sub-case, $\theta\in\QQ$) are not Diophantine, together
with a subclass of irrationals $\theta$, the \emph{Liouville
numbers} -- not satisfying \eqref{DiophantineCond}. In both cases,
SFE fails and quits to be useful for investigations, whose
achievement shall necessarily rely on other tools. Therefore one
summarizes the previous concepts as follows:

$\bullet$ when SFE commutes in a sufficiently close neighborhood
of (super--)attracting, indifferent and repelling points, the
local behavior of iterates links to only one among the three
elementary motions (either \emph{contraction}, \emph{rotation},
\emph{repulsion} respectively) [\footnote{The fourth linear
motion, \emph{translation}, refers back to contraction or
repulsion cases, depending on the vector direction approaching or
leaving $\delta$ respectively.}] with regard to the fixed point
$\delta$; elementary motions also include the translation
$f(z):z+k$ and the identity map $f(z):z$, known to apply at the
fixed points.

$\bullet$ so SFE classifies orbits dynamics by \emph{contraction},
\emph{rotation}, \emph{repulsion} exclusively; it turns out that,
as SFE fails, local dynamics do not involve only one elementary
motions, as explained in diagram \ref{MotionTable}: then
\textbf{Fatou-Leau flowers} and \textbf{Hedgehogs} consist of
compositions of more than one elementary motion. Hence \emph{the
failure of SFE shuts the door to linearizability.}

\subsection{Hedgehogs, the next stop}\label{Hedgehogs}
Stepping in the second sub-level of table \ref{IndifferentTable},
we assume to deal with holomorphic quadratic germ\footnote{The
current result apply to this hypothesis. For higher degree germs,
they are subjected to changes.}
\begin{equation}\label{Eq3}
f(z):e^{2\pi i\theta}z+z^2+\dots\ .
\end{equation}

\noindent Iterates $f_n(z)$ can be thus expressed in this form:
\begin{equation}\label{Eq4}
f_n(z):e^{2\pi i\theta n}z+\mathcal{O}(z^k),\qquad n\geq 1, k\geq
2
\end{equation}

\begin{figure}[h]
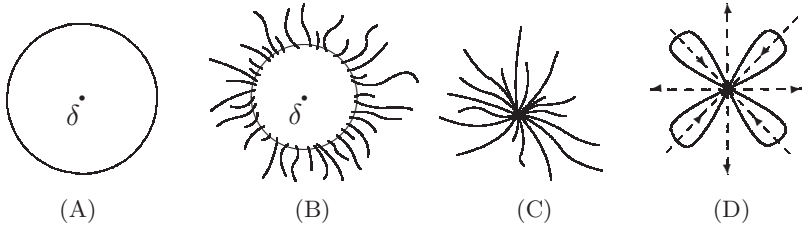

\centering
\begin{tabular}{cccc}
  \input{siegel.pic} & \input{hedge01.pic} & \input{hedge02.pic} & \rotatebox{5}{\input{flower03.pic}}\\
  \footnotesize (A) &\footnotesize (B) &\footnotesize (C) &\footnotesize (D)\\
\end{tabular}
\setcaptionwidth{0.9\textwidth}\captionstyle{normal}\caption{\textbf{The
ideal journey}. Simple figures of the local invariant sets related
to the numerical properties of the angle $\theta$ for indifferent
points: from the Siegel disc (A) to a Fatou-Leau flower
(D).}\label{Travel01}
\end{figure}

If compared to \eqref{SFE_Siegel}, the question boils down to the
study of the $\theta$ values so that $\mathcal{O}(z^k)$ vanishes
or not, together with the related topological dynamics. It is
straight-forward that $\mathcal{O}(z^k)=0$ allows to conjugate
locally the iterations to the rigid and aperiodic rotation
$f_n(z):e^{2\pi i\theta n}z$ (fig. \ref{Travel01}/A) in a
sufficiently close neighborhood of $\delta$. Topologically
speaking, if $\theta$ enjoys the diophantine condition
\eqref{DiophantineCond}, the invariant set is a disc-shaped
neighborhood, the \emph{Siegel disc} $\mathcal{S}$ centered at
$\delta$, which Julia termed \emph{center} like Poincar\'{e} did
for the analogous points in the field of differential equations:

\begin{definition}[Siegel Disc]\label{DefSiegelDisc}
Let $f$ be a non-linear complex function \eqref{Eq3} and let
$\delta$ be an irrationally indifferent fixed point, so that
$|f'(\delta)|=e^{2\pi i\theta}$. If $\theta$ enjoys the
diophantine condition \eqref{DiophantineCond}, the SFE
\eqref{SFE_Siegel} commutes and there exists a sufficiently close
neighborhood $\mathcal{S}$ of $\delta$, so that $\mathcal{S}$ is a
simply connected component of the basin and analytically
conjugated to an aperiodic rotation $e^{2\pi i\theta}$ of the unit
disc $\DD$. \textnormal{(\cite{Milnor}, p. 117)}
\end{definition}

\begin{figure}[h]
\centering \vspace{0.5cm}
\includegraphics[width=6.0cm]{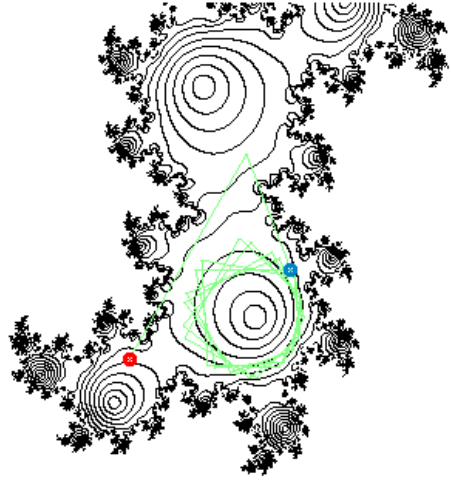}
\setcaptionwidth{0.8\textwidth}\captionstyle{normal}\caption{\textbf{Singular
dynamics}. Tracking down one orbit inside the basin of attraction
with a Siegel disc $\mathcal{S}$. The red and blue points mark the
seed $z$ and the final iterated image point $z_n$ respectively.
This figure was plot by nested circles
equi-potentials.}\label{Table11} \vspace{0.5cm}
\end{figure}

What happens if this condition is no longer met by
$\theta\in\RR\backslash\QQ$? Recent Mathematics has been stressing
that the concept of `\emph{Natura non facit
saltus}'\footnote{Ancient Roman phrasing by Linneus, alluding to
Nature evolution by gradual and infinitesimal changes.} rules out
for dynamical systems. Researches
(\cite{Cherry-1964,PerezMarco-1997,Yoccoz-1988,Yoccoz-1994}, to
quote a few) showed the existence of irrationals $\theta$
\emph{dropping gradually off} the diophantine condition
\eqref{DiophantineCond}: the Siegel disc $\mathcal{S}$ slowly
turns into a new topological family of invariant sets, namely the
hedgehogs $\mathcal{H}$. We cannot expect immediate changes in the
local geometry: $\mathcal{S}$ squeezes into a smaller disc and
neighboring orbits inside $\mathcal{B}\backslash\mathcal{S}$ get
very complicated; if $\mathcal{S}$ is not maximal, we say that
$\mathcal{H}$ has a small linearization area. In fact, with regard
of the angle value $\theta$ involved, $\mathcal{S}$ may shrink to
the fixed point $\delta$: here $\mathcal{H}$ has no linearization
area and $\delta$ is defined a \emph{Cremer point}\footnote{In
honor of Hubert Cremer (1897--1983) who first proved their
existence when $\theta$ does not meet the Diophantine condition.}.

During 1920s and 30s, Cremer showed \cite{Cremer-1927} that the
set $\RR\backslash\QQ$ of all irrationals can be partitioned into
two disjoint subsets, $\mathcal{M}, \mathcal{N}$:

$\bullet$ let $\mathcal{M}$ be the set of angles $\theta$ so that
$\mathcal{D}_c$ is enjoyed, then a maximal Siegel disc
$\mathcal{S}$ exists; $\mathcal{M}$ is a full Lebesgue's measure
set;

$\bullet$ let $\zeta\in\mathcal{N}$ be angles whose value is a
Liouville number. $\mathcal{N}$ is a null-measure set.

In the course of his proof on the existence of non-centers for
rational maps \cite{Cremer-1927}, Cremer also showed that the
iterates $R^n(z)$ give rise to periodic cycles of growing order
$n$ in the neighborhood of the non-linearizable indifferent
$\delta$. He also found that they accumulate at $\delta$ while
\begin{equation}
\label{CremerProblemEq16} \liminf_{n=1,2,\dots}
\sqrt[s^n]{\abs{\beta^n-1}}=0
\end{equation}

\noindent which holds for
\begin{equation}
\label{CremerProblemEq17} \liminf_{n=1,2,\dots} \abs{\beta^n-1}=0
\qquad\textnormal{\footnotesize\em and}\qquad
\liminf_{n=1,2,\dots} \beta^n=1, \beta=e^{2\pi i\theta}.
\end{equation}

\begin{figure}[h]
\centering
\begin{tabular}{cc}
  \includegraphics[height=3.5cm]{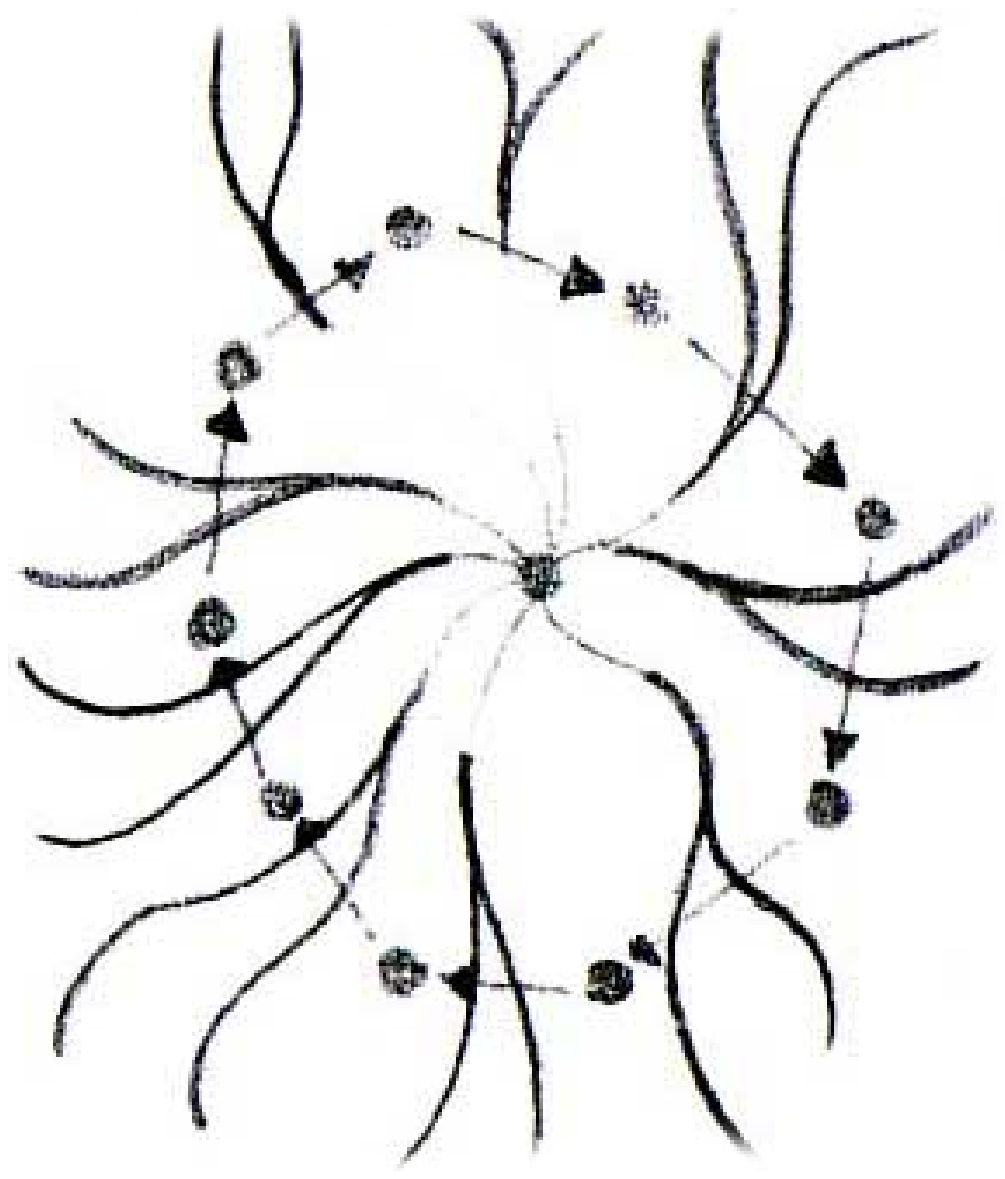}&
  \includegraphics[height=3.5cm]{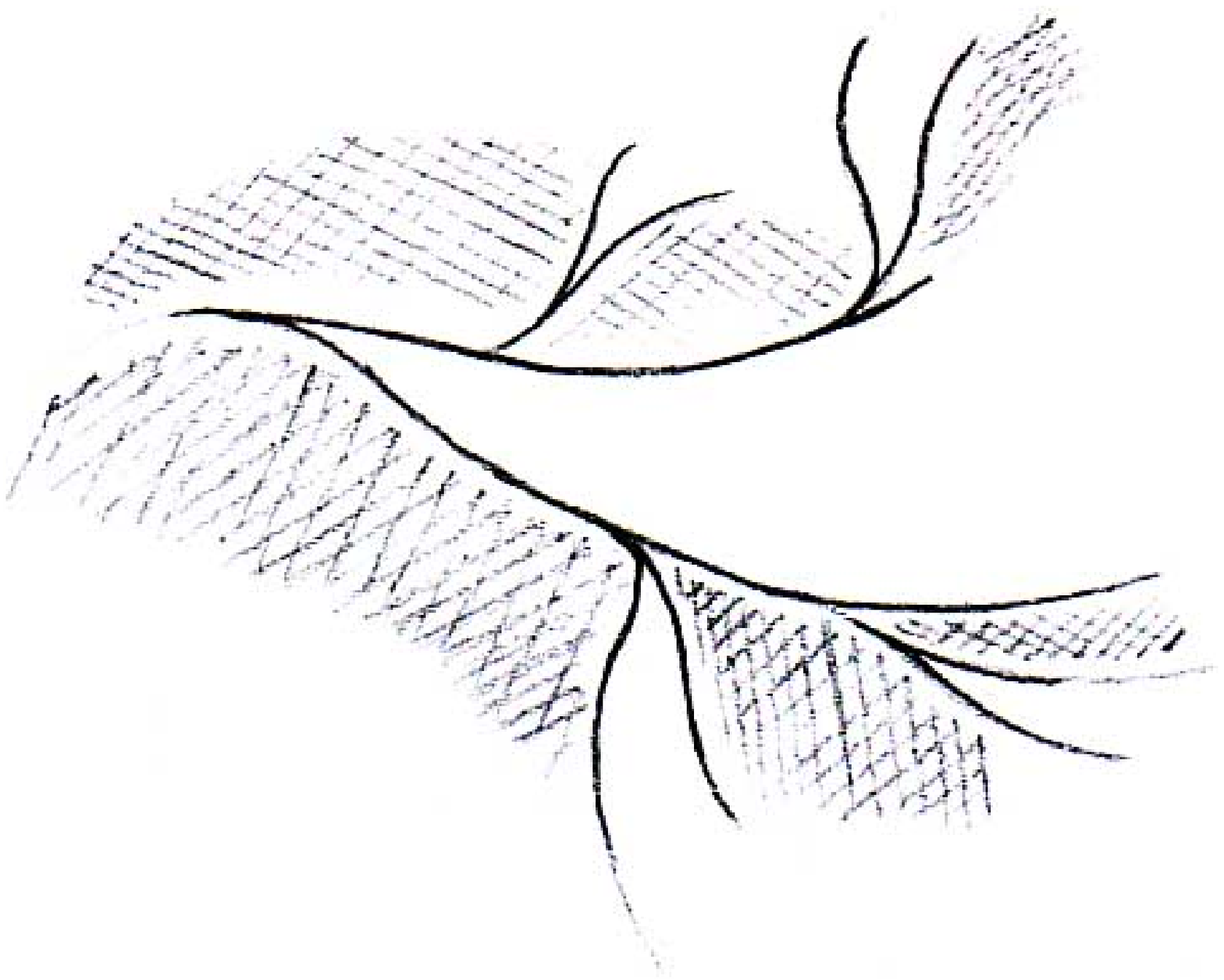}
  \\
  \footnotesize (A) &\footnotesize (B)\\
\end{tabular}
\setcaptionwidth{0.9\textwidth}\captionstyle{normal}\caption{\textbf{Wedging
the bounded basin (I)} Two hand-made drawings of the hedgehog
entering the bounded basin (shaded region). The wedging white
region is the basin of the point at infinity.}\label{Hedge45-46}
\end{figure}

\begin{figure}[h]
\centering
  \input{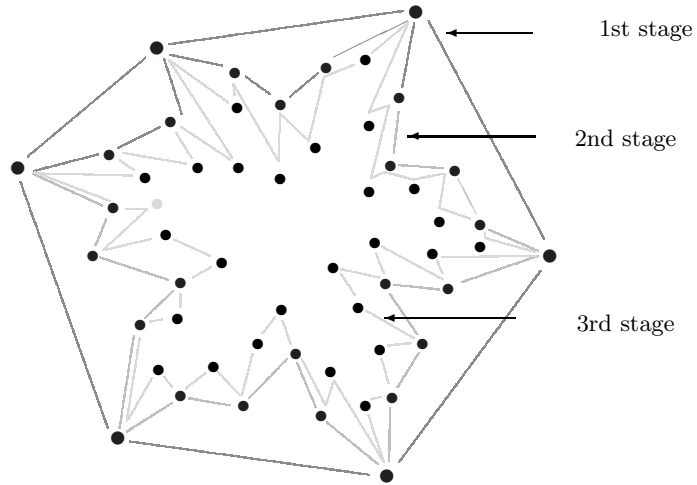}\\
\setcaptionwidth{0.7\textwidth}\captionstyle{normal}
\caption{\textbf{Hedgehogs generation.} An illustration of $3$
steps (different shades of grey) of repelling cycles accumulation.
The starred-shape, due to the cycles getting near $\delta$, as
well as wedging action of the Julia set $J$, are
evident.}\label{HedgeDyns}
\end{figure}

\noindent Since the number of (super--)attracting cycles is finite
(according to Koenigs' conjecture in 1883 and proved to hold by
Fatou and Julia) and the Julia set $J$ nature (closure of
repelling cycles), most of the periodic points in these cycles
have to be repelling\footnote{Cremer never questioned on the
nature of the accumulating cycles at $\delta$ throughout his
works.} so that, roughly speaking, \cite{Cherry-1964} their
accumulation shows that, as $n$ grows, $J$ wedges (see figs.
\ref{HedgeDyns} and \ref{Hedge38-40}) the bounded basin of
attraction: if $n$ has no upper bound, then $\delta\in J$,
otherwise the wedging action of $J$ stops at a certain distance
from $\delta$ for $n\rightarrow K<+\infty$. The zero limit in
\eqref{CremerProblemEq16} is reached when $\alpha$ is
rational\footnote{This case is out of our context. Anyway the
Fatou-Leau flower dynamics are inductively helpful to watch how
the basin to $\infty$ wedges the other basin, so that the Julia
sets attaches to $\delta$, assumed the germ $f(z):e^{2\pi
i\theta}z+z^n$, where $n\in\NN^+,n\rightarrow\infty$ and
$\theta\in\QQ$.} or it is a Liouville number; hence a poorly
rational approximation -- for the \index{Diophantine
condition}diophantine condition for $\theta$ -- implies rotation
around $\delta$, because cycles do not accumulate at it.
Inspired by continuity, it is suggestive to think of arranging an
ideal journey, along the number properties so to visit all
invariant sets arising for indifferent fixed point, from (A) to
(D), that is, numerically speaking in terms of $\theta$, from
$\RR\backslash\QQ$ to $\QQ$. During the journey, the Siegel disc
$\mathcal{S}$ squeezes (fig. \ref{Travel01}/B) until it disappears
(fig. \ref{Travel01}/C); for both cases B and C, the invariant
sets are `hedgehogs' $\mathcal{H}$, when $\mathcal{S}$ is not
maximal or has empty interior. This additional look-up at the
journey is inspired by the strong unifying power offered by
hedgehogs theory, which links apparently far dynamical
configurations. The mathematical definition of the hedgehog
$\mathcal{H}$ is \cite{PerezMarco-1997} as follows:
\begin{definition}[Hedgehog]
Given a neighborhood $U $of an irrationally indifferent point
$\delta$, so that the holomorphic map $f$ is univalent on $U$, the
hedgehog $\mathcal{H} is an invariant compactum\footnote{A
compactum (plural, compacta) is a compact metric space.}$, so that
$f$ is not linearizable or has linearization domain relatively
compact in $U$.
\end{definition}

For \eqref{Eq3}, the current research investigates on the
semi--local dynamics about $\mathcal{H}$ and thanks to the results
by P\'{e}rez-Marco in 1990 (\cite{Milnor}, p. 123) one knows that
$\mathcal{H}$ may show up in these different three
characterizations:
\begin{enumerate}\label{HedgehogsList}
    \item hedgehogs $\mathcal{H}$ are locally linearizable and with no small cycles;
    \item hedgehogs $\mathcal{H}$ are not locally linearizable and with small cycles;
    \item hedgehogs $\mathcal{H}$ are not locally linearizable and without small cycles;
\end{enumerate}

The `\emph{small cycles}' property refers to the possible
existence of infinitely many cyclic orbits in a sufficiently small
neighborhood of the irrationally indifferent $\delta$. It is said
that `$\delta$ \emph{has the small cycle property}' or `\emph{is
approximated by small cycles}'. Other results on small cycles are
included in \cite{Yoccoz-1988,Yoccoz-1994}.

\section{Entering the computer graphics}\label{Entering}
\subsection{Features of ordinary methods}\label{OrdinaryMethods}
Most available methods are global and devoted to the display of
Julia sets $J$. One first notices that they do not
often\footnote{The approach by inverse maps achieves it, but it is
not exportable to any maps because inverse ones cannot be always
retrieved.} focus on the analytical properties of $J$. Both its
location and shape come out from the human eye perception of the
different colors distribution inside a close neighborhood of $J$.
In mathematical terms, these methods do not draw $J$: they paint
the Fatou set $\mathcal{F}\equiv\riemannsphere\backslash J$.
Therefore they are \emph{not detective}, but \emph{deductive}:
focusing on the complement set $\mathcal{F}$, they deduce $J$ from
it.

\subsection{Palette of colors}\label{subsubPalette}
Experience induces this consideration for color figures: the
(ascendent or descendent) ordered palette of shades\footnote{For
example, sorted by gradient: shades of blue or tones from green to
violet.} fits \emph{for understanding the convergent or
divergent\footnote{With regard to the nature of $\delta$.}
movement of iterates} between distant neighborhoods of the fixed
point, because the shades sequences play as graphical indicator of
the dynamics, but they fail to evince the local dynamics: the
convergence/divergence rate about the indifferent $\delta$ is very
slow and iterated domains match to very close colors in the
palette, looking like the same.

\subsection{Convergence criteria}\label{subsubConvergence}
The fate of iterated orbits is tested at each $f_n(z):z_n$, for
preventing infinite looping. This is commonly achieved by a
trapping disc $\mathcal{D}$ of given radius $r$; then
$\abs{z_n}<r$ is tested. This approach, aiming to understand when
(in terms of the value of the iterative index $i$) orbits might
escape $\mathcal{D}$, is commonly known as `\emph{escape time}'
algorithm. Some pseudo-code follows:

{ \footnotesize

\begin{verbatim}
    #include "complex.h" // this is a class handling complex numbers
                         // downloadable from author's site
    complex z, next_z ;
    complex c(0, -1);    // this is the complex parameter z = 0.0 - 1.0i

    t = 50 ;             // top iteration index to prevent infinite looping
    r = 2.0 ;            // the radius of the trapping disc

    for( int i = 0; i < t ; i++ )
    {
        next_z = z * z + c ;  // we assumed the quadratic iterator
        if ( abs( next_z ) > r ) break;

        z = next_z ;
    }

    take-some-color-value-from-the-iterative-index-or-from-the-point-z ;
    plot-z-on-the-screen ;

\end{verbatim}
}

Another approach relies on the convergence properties in the
neighborhood of $\delta$ and uses the distance between two
successive iterates as test condition:
\begin{equation}\label{Eq1}
\abs{f_i(z)-f_{i+1}(z)}<\epsilon,\qquad,\epsilon>0
\end{equation}

\noindent It is a variant of escape time and defined
the`\emph{approximation}' method:

{ \footnotesize
\begin{verbatim}
    #include "complex.h" // this is a class handling complex numbers
                         // downloadable from author's site
    complex z, next_z ;
    complex c(0, -1);    // this is the complex parameter z = 0.0 - 1.0i

    t = 50 ;             // top iteration index to prevent infinite looping
    e = 0.00001 ;        // the distance ranging between 0 < e < 1

    for( int i = 0; i < t ; i++ )
    {
        next_z = (2*z*z*z+1)/(3*z*z) ;    // this is the transformed map
                                          // of z^3-1 by Newton's method

        if ( abs( next_z - z ) < e ) break;
        z = next_z ;
    }

    take-some-color-value-from-the-iterative-index-or-from-the-point-z ;
    plot-z-on-the-screen ;

\end{verbatim}
}

\subsection{Obsolescence with local invariant sets}\label{Obsolescence}
\begin{figure}[h]
  \centering
\begin{tabular}{ccc}
  \includegraphics[width=2.5cm]{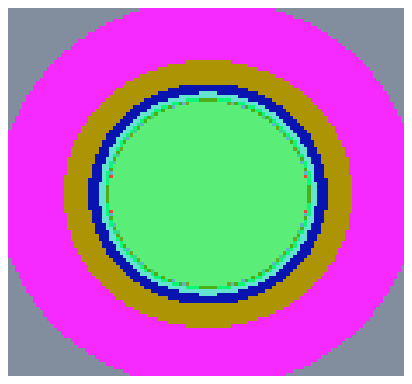} &
  \includegraphics[width=2.5cm]{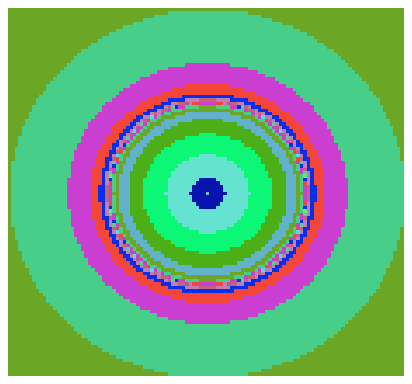} &
  \includegraphics[width=2.5cm]{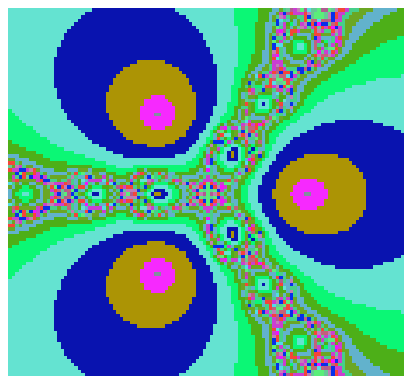} \\
  \footnotesize (A) &\footnotesize (B) &\footnotesize (C)\\
\end{tabular}
\setcaptionwidth{0.9\textwidth}
\captionstyle{normal}\caption{\textbf{Attempts of improvements}.
The random palette coloring for attracting fixed points. More than
the shading gradient approach, the attracting dynamics are evident
here from the nested discs shrinking up to points.}\label{Table01}
\end{figure}

We are going to discuss ways to bypass or lessen the two problems
in subsections \ref{subsubPalette} and \ref{subsubConvergence};
but even any improvement in such two directions would be still
insufficient for our goals. First we deal with colors: sequences
shades showed up to unfit, so might a randomly generated palette
help? In the color figures of table \ref{Table01}, we iterated the
neighborhoods about fixed points of different nature and painted
them by the random palette. For figs. \ref{Table01}/A and B, we
picked up the square map $f(z):z^2$, having a super--attracting
fixed point $\delta$ at $0$; in (C), the Newton-Raphson method was
applied to $f(z):z^3-1$, having $3$ attracting points on the unit
circle $\partial\DD$.
One notices that this works finely with the neighboring dynamics,
whose shrinking behavior to the attracting fixed point(s) is
displayed step by step, or analogously in the neighborhood of a
repelling point.

\begin{figure}[h]
  \centering
\begin{tabular}{cccc}
  \includegraphics[width=2.5cm]{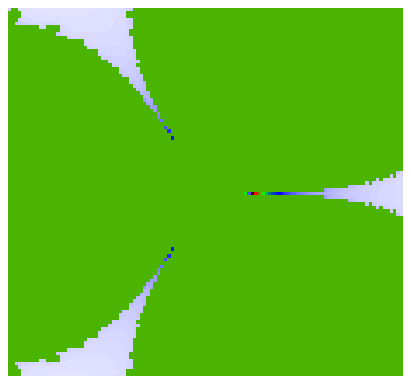} &
  \includegraphics[width=2.5cm]{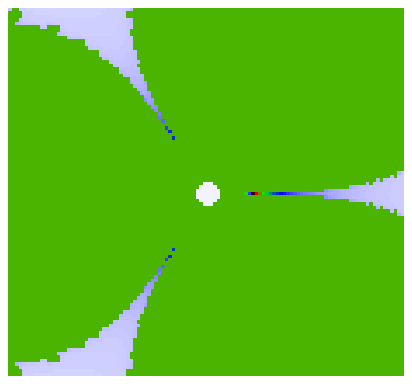} &
  \includegraphics[width=2.5cm]{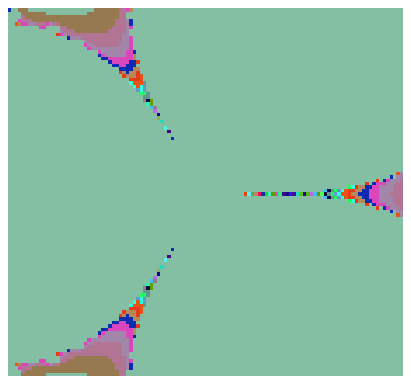} &
  \includegraphics[width=2.5cm]{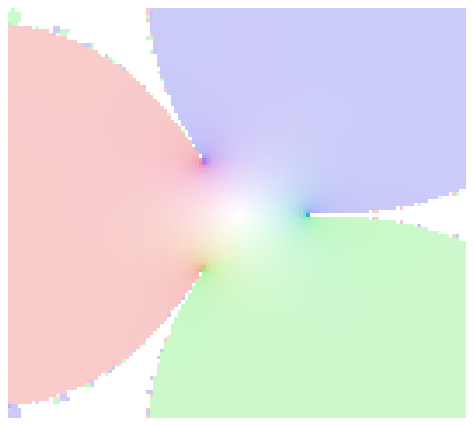} \\
  \footnotesize (A) &\footnotesize (B) &\footnotesize (C)&\footnotesize (D)\\
\end{tabular}
\setcaptionwidth{0.9\textwidth}
\captionstyle{normal}\caption{\textbf{Ordinary attacks to
rationally \dots\ }. Different methods for the Fatou--Leau
flower.}\label{Table02}
\end{figure}

\begin{figure}[h]
  \centering
\begin{tabular}{cccc}
  \includegraphics[width=2.5cm]{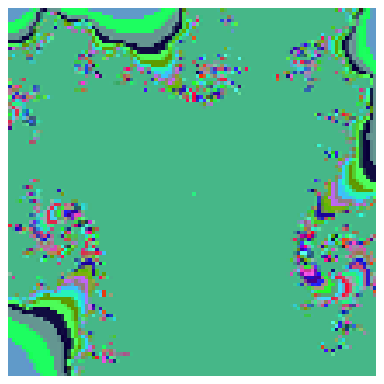} &
  \includegraphics[width=2.5cm]{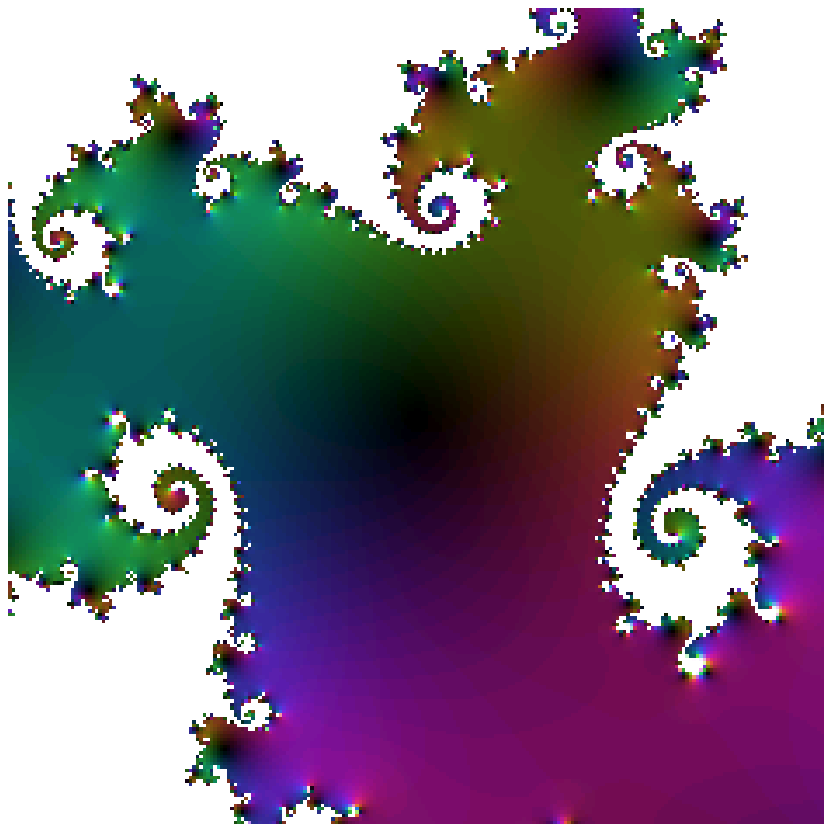} &
  \includegraphics[width=2.5cm]{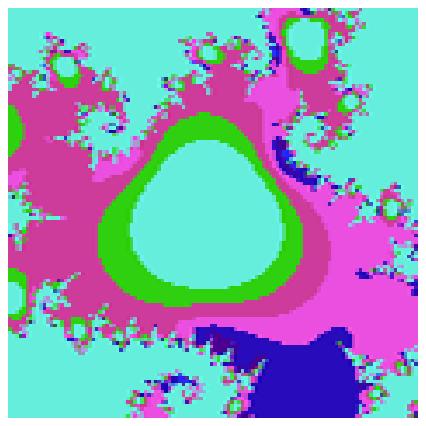} &
  \includegraphics[width=2.5cm]{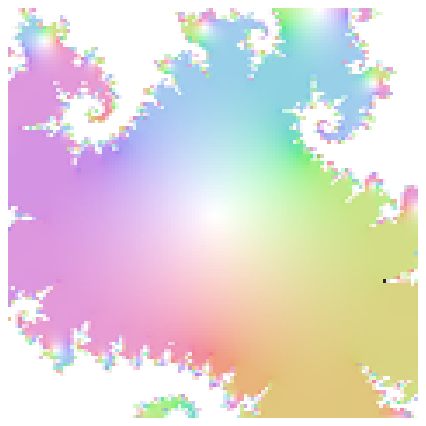} \\
  \footnotesize (A) &\footnotesize (B)&\footnotesize (C)&\footnotesize (D)\\
\end{tabular}
\setcaptionwidth{0.9\textwidth}
\captionstyle{normal}\caption{\textbf{\dots and irrationally
neutral points}. Dynamics near a Siegel disc (A), (B) and an
hedgehog (C), (D).}\label{Table03}
\end{figure}

In the figures of tables \ref{Table02} and \ref{Table03}, we
painted all invariant sets (refer to fig. \ref{Travel01}) which
could possibly arise in the neighborhood of indifferent points
$\delta$ by ordinary method, with emphasis to colors sequence and
values accuracy.

The iterations of $f(z):z+z^4$ yield a `Fatou-Leau flower' with
$3$ petals meeting all together at the origin. Either with the aid
of random palettes or as infinitesimal values of $\epsilon$ are
set into inequality \eqref{Eq1}, one cannot evince the local
dynamics, but only the basins shape. And no benefit is drawn from
the experiments with irrationally indifferent points. In fig.
\ref{Table03}/A and B, the random palette and a very high number
of iterates were respectively tested to work with a Siegel disc
case (it should appear in the green basin). The plot of an
hedgehog was tested in (C, D). In particular, (B) and (D) use
another coloring method which sets a one-to-one map between each
point and the RGB cube, so that any complex point is univocally
colored with regard to his location. Even if the iterative index
$i$ increases to very huge values, orbits cannot reach too close
to $\delta$ so we cannot get better pictures of it.

\section{Off to `composite dynamics'}
\subsection{The core questions}\label{CoreQuestions}
What does to draw a local invariant set mean? And, in a larger
sense, what do we want to get while drawing it?

Without basing upon good responses to these questions, it does not
make so much sense to go on with our attempt. As remarked in table
\ref{MotionTable} at p. \pageref{MotionTable}, holomorphic
dynamics have different invariant sets.

Julia sets $J$ are easy to display: the only issues are merely
technical (methods, palette colors, \dots). In local terms, the
neighboring dynamics about points of $J$ can be easily collected
under one easy concept: orbits, whose character is initially
repelling near $J$, run across the basins $\mathcal{B}$ of
attraction up to a (super--)attracting cycle (here, a fixed point
$\delta$).

Some local invariant sets $\mathcal{I}_s$ around $\delta$ enjoy
more or less complicated dynamics. The table \ref{MotionTable}
offered an overview of cases, where we noticed invariant sets
endowed with `\emph{composite dynamics}', that is, involving more
than one elementary motion: this is a bunch of information which
is harder to be pulled out than when SFE commutes and there is
only one local elementary motion to deal with. One should also
take in account that the convergence speed of neighboring orbits
may get slower as they reach closer to $\delta$: in this case,
understanding such motions via colors is impracticable. Here one
needs to evince either the (1) boundary $\partial\mathcal{I}_s$
(retrieving the `shape') and the (2) inner motions (figuring out
the `behavior' of orbits).

Reaching (2) needs to consider that the elementary psychology of
Space relates the perception of motion `\emph{to lines in the
principal directions of curvature, which may communicate surface
shape better than lines in other directions}'
(\cite{GiInHaLe-1986}, p. 43). This is a common feature of the
field of Non-Photorealistic Rendering (NPR) and associated to the
concept of `\emph{perceptually efficient images}', which is a
visual representation emphasizing important features and
minimizing superfluous details: our rendering are somehow skinny.
For our context, we find useful to filter this concept by our
needs: we want an analogously-shaped geometrical model of the
given vector field, which is capable of evincing (1) and (2).

\subsection{Obstructions (I) : no ticket for the ideal journey}\label{ObstructionsI}
The problem of drawing hedgehogs on a computer was first posed to
the author by professor P\'{e}rez-Marco during an informal meeting
in summer 2002. In general, the degeneration of a value
$\theta\in\RR\backslash\QQ$ into $\theta\in\QQ$ by \emph{finite
digits machine computation} is quite obvious.

Assuming the formula \eqref{Eq4}, one major issue is to check
whether the resulting $\zeta=\theta^n$ may approximate irrationals
poorly or not. One would like to distinguish the formal value
$\theta$ with the input one $\alpha$ in practice, where
$\alpha\approx\theta$ necessarily holds: here
$|\alpha-\theta|=\varepsilon>0$. Analogously, one has
$|\alpha^n-\theta^n|=\rho>\varepsilon>0$ for iterates, thus the
error magnitude grows with the iterative index $n$. Resuming, we
are not contented by the value $\theta$: how could we reduce
$\rho$, if possible? In fact subtle issues may come up about the
possibility: first, Cremer's partition of irrational numbers shows
that values of the null-measure set $\mathcal{N}$ are almost
impossible to pick up by a machine with arbitrary precision, in
order to guarantee the chosen invariant set: for example,
hedgehogs $\mathcal{H}$ with no linearization area relate to
extremely weak numerical conditions under the decimals cut-off, so
that one could never watch them by relying on numerical
computations. \emph{So would it make sense to approximate a
number} $\alpha^n\in\RR\backslash\QQ$ \emph{via another
irrational} $\theta^n$ \emph{which possibly enjoys different
numerical properties?} We reply that the value $\alpha^n$ may give
rise to a local invariant set differing, more or less evidently,
from what we really wanted from $\theta^n$. We observe that from
Lebesgue null-measurability, it does not follows that
$\mathcal{N}$ should be pointwise and totally disconnected; rather
it consists of disjoint intervals whose amplitude is positive but
infinitesimal, say, of width $\varepsilon$. \emph{If machine
approximation may actually work under such interval}
$\varepsilon$\emph{, we might think of being able to get very good
approximation of the original value} $\alpha^n$ \emph{and of
finding the related invariant set with adequate precision.}

Without this chance, the approximation problem is weakly
attainable on a numerical basis; is there an alternative attack
relying on the approximation of properties? We recall that
approximation in itself refers to another entity $\beta$, inexact
and imitating (not copying) the original $\alpha$; so, dealing
with properties, we questioned on whether it is plausible to seek
one property relating to a third value $\beta$ and so that
$$\abs{\alpha - \beta}<\abs{\alpha - \theta}$$
holds? We would enter an \emph{ad infinitum} process generating an
infinitely many values $\xi,\gamma,\dots,\tau,$ and so that
$$\abs{\alpha - \beta}<\abs{\alpha - \xi}<\abs{\alpha - \gamma}<\dots <\abs{\alpha - \tau}<\abs{\alpha - \theta}$$
and therefore infinitely many properties $\mathcal{P}$ standing
between the formal value $\alpha^n$ and the practical value
$\theta^n$ and even nested as follows:
$$\mathcal{P}_\beta\subset\mathcal{P}_\xi\subset\mathcal{P}_\gamma\subset\ \dots\ \subset\mathcal{P}_\tau\subset\mathcal{P}_\theta\ .$$
\begin{question}\label{ObstructionsIQuestion01}
Is the number of numerical properties finite? If so, how many they
are?
\end{question}
We conjecture that the set of properties $\mathcal{P}$, involved
in the local linearization around an irrationally indifferent
$\delta$, is finite: because the topological configurations of
hedgehogs are also finite.

\subsection{Obstructions (II): practice is what matters}\label{ObstructionsII}
Now with regard to the practical computations involved in the
iteration process, we can find three classes of obstructions:
\begin{enumerate}
    \item \emph{Statistical}: due to the $\mathcal{M},\mathcal{N}$ sets distribution over the real interval $[0,1]$,
    it much easier to pick up a $\theta\in\mathcal{M}$ than one
    $\zeta\in\mathcal{N}$. And due to above approximation, this task gets
    harder than ever.\label{Obstruction2}
    \item \emph{Numerical}: numbers $\theta\in\mathcal{M}$ are more
    robust to  such approximation attack than $\zeta\in\mathcal{N}$;
    in fact, under iterates, Liouville numbers $\zeta$ tend to be turned into new values
    which are Diophantine irrationals; new resulting values cannot goodly approximate rational numbers
    and then one falls back into the Siegel disc case.\label{Obstruction1}
    \item \emph{Procedural}: Liouville's condition is fundamental to plot
    hedgehogs, but it is also weak to be preserved under iteration.
    The movement rate or orbits in the hedgehog $\mathcal{H}$ gets
    as slower as closer to the fixed $\delta$. Thus one would be pushed to increase
    the iteration index $n$ to get finer results but we saw the iteration process supports
    obstructions \ref{Obstruction2} and \ref{Obstruction1}.\label{Obstruction3}
\end{enumerate}

Approximation and computer usage are two `obstructions' we cannot
overcome. Not having a valid solution at hand nor being successful
to find it, the problem was left open until we occasionally came
to it during March 2006, when we planned to find a strategy for
\emph{lessening the break of irrationality}. It was clear that the
numerical path was impracticable.

Our problem with hedgehogs refers back to the impossibility of
drawing local invariant sets, where slower and slower convergence
rates, as iterates are closer to $\delta$, escape the
possibilities offered by the standard methods: in the next
section, an analysis on the ordinary graphical methods for
holomorphic dynamics will be given in order to motivate their
exclusion from the run for displaying local invariant sets and,
most important, hedgehogs accurately.

\subsection{Quality vs. quantity}\label{OffTo}
The methods described in section \ref{Entering} are
\emph{structurally weak} for our purposes: the reader shall know
that they have been developed (read, customized) for iterates of
polynomial maps, in primis the quadratic $f(z):z^2+c$, where
$z,c\in\CC$ and $c$ is a parameter. As we showed here and even for
quaternionic Julia sets, as discussed by the author in
\cite{Rosa-2005,Rosa-2006}, they might not work finely or even
they may fail completely if exported to dynamical systems which
are distinct from their original context. We remark that, mostly due
to the very slow convergence rate of orbits near the indifferent
$\delta$, the escape time approach is inconvenient to display the
hedgehogs, either if different colors sets are applied or if the
index $i$ is hugely increased. Ordinary methods work
\emph{quantitatively} and they fail with dynamics wanting the sharpest
numerical precision; in addition, the geometry of these
invariant sets (both in local and global terms) is exclusively
retrieved by the value of the last iterated point $z_m$, where $m$
is any largest iterative index $i$ available to the machine
architecture. We may set $m$ to the largest value then, but
computations would be excessively time consuming.

In the spirit of automatism helping to save time, we state that a
radically different method is required: thus a way out is not to
reach to those neighborhoods, but to
\label{be-already-there-and-work}`\emph{be-already-there-and-work}'.
Even in this new direction, the method customization is required:
in fact, we judged useful to adopt an half-way strategy which,
besides the inevitable finite digits computation, teams up with a
\emph{qualitative} attack by \emph{the imitation of a model with
pre-defined shape}.

The decision on the model matured after the conclusions
in section \ref{AnalogyModels}, where the failure of SFE for
Fatou-Leau flowers and hedgehogs implied that such dynamics cannot
be adequately described by models based upon regular curves, such
as concentric circles or sheaves of straight-lines. So we wanted
to adopt a non--regular and, mostly, \emph{topologically equivalent model}.
Namely, this is the (holed) $n^{th}$-branched star, as depicted in
table \ref{Table04}. It is extremely important to remark that the
following results are based upon the iteration of quadratic germ \eqref{Eq3}.
Actually there exist obstructions of theoretical nature, stopping to
extend the results from quadratic holomorphic germs to higher degrees.

\section{Re-elaborating the equi-potentials}\label{EquiPotentials}
\subsection{The holed $n$-th branched star model}\label{HoledBranchedStar}
Our approach roots to nothing new in digital graphics for
dynamical systems and often used in complex analysis. Known under the term of `\emph{equi-potential curves}',
the reader can find examples of it in \cite{Needham-2000} by Needham
and, for holomorphic dynamics, in \cite{Milnor} by Milnor --
although it already appeared, but weakly, in older publications
\cite{PeRiSa-1991, PeSa-1988}. The further pseudo--code, reported to
support the concept rather than being implemented as it is, consists of a main
routine assuming the whole input finite subregion of $\CC$ as a grid
of seeds $z$ and filtering the dynamics of each iterated point $z_n$
via a sub-routine checking if an orbit can be possibly plot, according
to the holed star model.

{ \footnotesize
\begin{verbatim}
    #include "complex.h"  // this is a class handling complex numbers
                          // downloadable from author's site

    complex fxd_pt ;      // we assume that we already
    fxd_pt.real = 0.0;    // know the location of the fixed point
    fxd_pt.imag = 0.0 ;   // around which we want to plot the hedgehog

    double val = 0, tmp_val = 0 ;   // values which will be later
                                    // stored and compared
    BOOL bInitFlag = FALSE ;        // this boolean flag will be
                                    // to check the two above [double] values

    // we assume that the four coordinates of the screen port where
    // the image will appear are stored in these variables;

    int top = 0, bottom = 320, left = 0, right = 200 ;
    complex z ;

    for ( x = left + 1; x < right ; x++ )
    {
        for ( y = top + 1; y < bottom ; y++ )
        {
            // suggestion: do a converse estimation map to come from screen
            // coordinates to the complex point z further by 'RescaleToMap'
            // so that you're assured that decimals approximation
            // does not make more than one screen point to the same
            // complex point z

            rescale-the-pair ( x, y ) to-the-pair ( z.real, z.imag );
            ////////////////////////////////////////////////////////////

            // at this point you shall just iterate the function
            // for the required number of steps but without
            // test conditions. See previous code.
            // Just stop it by a iterative top index limiter.

            // We did not show any related code for the function
            // to be iterated because one needs more complicate
            // explanations.
            // Make sure you are working with a suitable function
            // retrieving an hedgehog. We suggest you to use the one
            // an holomorphic quadratic and indifferent germ where
            // the angle theta is represented by the continuous fraction
            // reported further.

            // We can say we coded a complex parser
            // to input any function even in the form; sorry
            // but the latex syntax is the best we can use to
            // arrange one example:
            // f(z) : e^(2\pi i\theta)z+z^2

            complex output_z = iterate-the-function f(z) ;

            // NOTE : if you input the identity map, you'll see
            // the holed branched star.

            BOOL bDraw = Holed_Star( output_z, fxd_pt, double& val,
                                     double& tmp_val, BOOL& bInitFlag ) ;

            if ( bInitFlag )
            {
                if ( val != tmp_val )
                {
                    if ( bDraw ) pDC->SetPixel( x, y, 0 );
                    val = tmp_val ;
                }
            }
            else bInitFlag = TRUE ;
        }
    }
\end{verbatim}
}

In the previous code, one sees a \emph{raster\footnote{This is a
technical term in computer science to indicate that an image is
regarded as a mesh of points distributed in rows and columns; each
point is associated to a triplet of values. The first two values
are the unique pair of coordinates which define the location in
the mesh. The third value refers to the point color, usually
defined in the RGB additive model.} scan} of the screen by rows
(\texttt{y}) and columns (\texttt{x}). As integer coordinates are
taken per each screen point, they have been immediately turned
into real and imaginary values respectively, so to have finally
the complex point \texttt{z}.

\begin{figure}
  \centering
\begin{tabular}{cccc}
  \includegraphics[width=2.5cm]{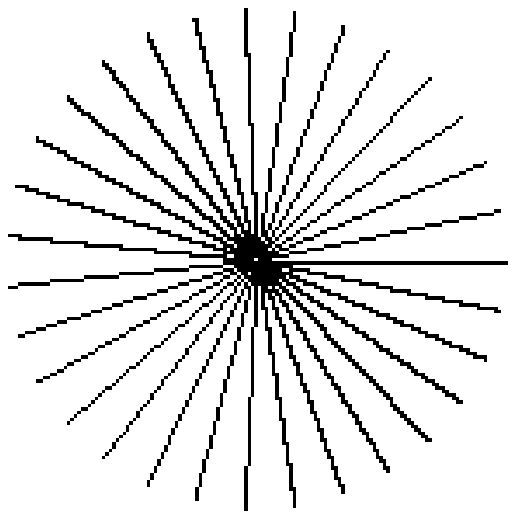} &
  \includegraphics[width=2.5cm]{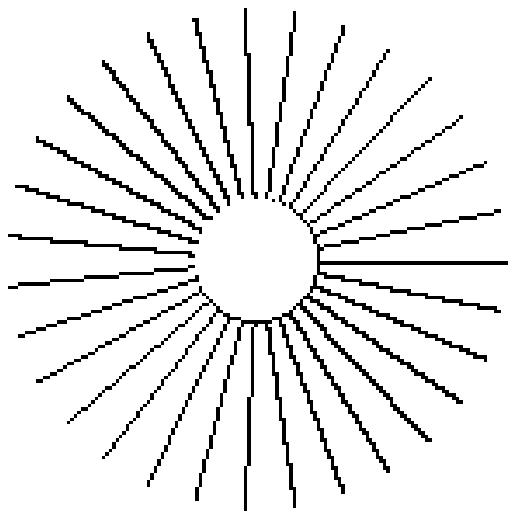} &
  \includegraphics[width=2.5cm]{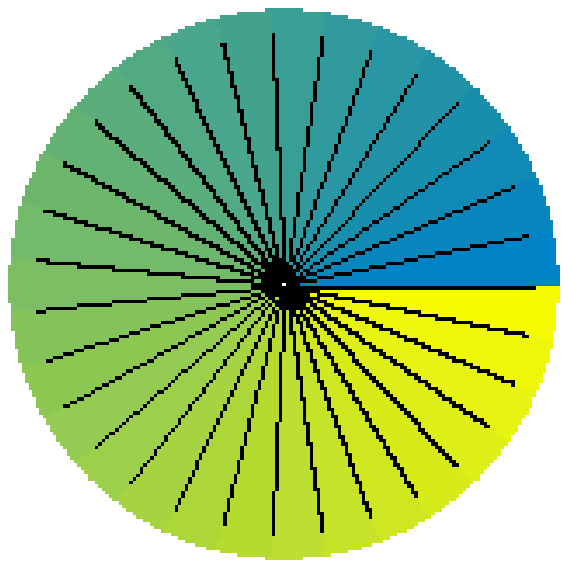} &
  \includegraphics[width=2.5cm]{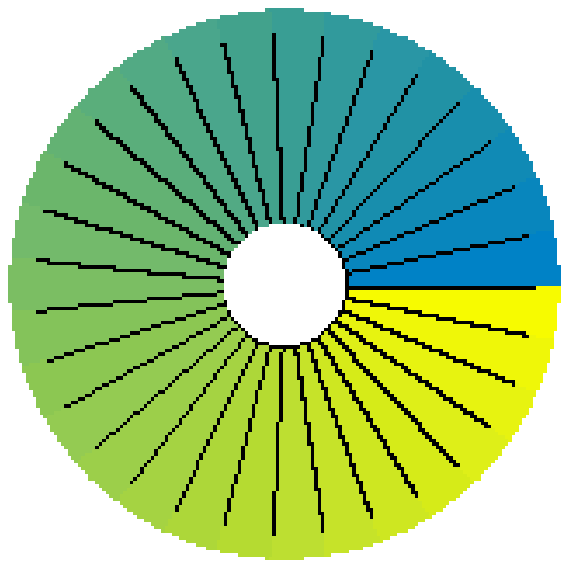} \\
  \footnotesize (A) &\footnotesize (B) &\footnotesize (C)&\footnotesize (D) \\
\end{tabular}
\setcaptionwidth{0.9\textwidth}
\captionstyle{normal}\caption{\textbf{I am the star}. The (holed)
$n^{th}$-branched star in monochromatic and colored
versions.}\label{Table04}
\end{figure}

And now we give the code details for the function
\texttt{Holed\_Star} which draws the hedgehogs/holed star by
equi-potentials point after point. In this code, we just presented
monochromatic hedgehog (refer to table \ref{Table04}, A/B), the
colored version requires one more function which we chose to not
include here for code simplicity.

{ \footnotesize

\begin{verbatim}
BOOL Holed_Star( complex z, complex fxd_pt, double& val,
                 double& tmp_val, BOOL& bInitFlag )

{
    #include "complex.h" // this is a class handling complex numbers
                         // downloadable from author's site

    #define PI 3.14159265358979323846  // we need PI to be as
                                       // precise as possible

    complex tmp_z = z ;         // we need another variable for
                                // storing a temporary value later

    double branches_number = 12 ;      // this explains itself !
    double offset = 0.5 ;              // this is the radius of the
                                       // disc hole in the star
                                       // if set to zero, the disc is empty

    double existence_interval = 2.0 * PI ;   // the values range in which
                                             // compute all branches location
                                             // in radial terms

    double potential_rate = existence_interval / branches_number ;

    tmp_z -= fxd_pt ;            // performs a translation of all points
                                 // so that the fixed point maps to the
                                 // origin and the further computations
                                 // get easier
    double v = tmp_z.angle();

    double out_level = ( f.euclidean_dist( z ) <= offset ) ? -1 : v / potential_rate ;

    int ol = (int)out_level ;       // we need an integer value for the
                                    // output level to be eventually used
                                    // for color the point

    if ( !bInitFlag ) val = ol ;    // if the flag was not set,
                                    // then this means that it is the first ever
                                    // value to be stored.
    else tmp_val = ol ;             // Otherwise, it does not. See before
                                    // in the main function

    if( ol < 0 || ol > branches_number )     // check for values not
                return FALSE ;               // ranging out of the interval

    return TRUE ;
}
\end{verbatim}
}

We have now a bunch of code to scan a region
$\mathcal{R}\subset\CC$, so that \texttt{Holed\_Star} classifies
the $n$--fold image region $f_n(\mathcal{R})$ via the star model.

\begin{figure}[h]
\centering
\begin{tabular}{ccc}
  \includegraphics[width=3.8cm]{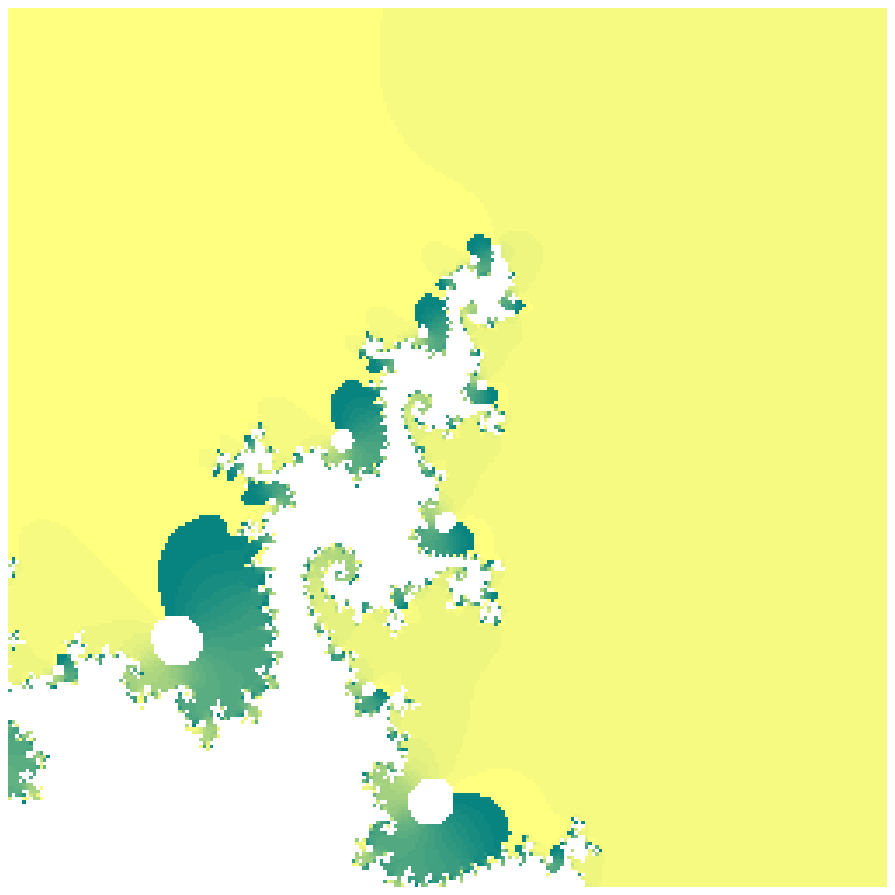}&
  \includegraphics[width=3.8cm]{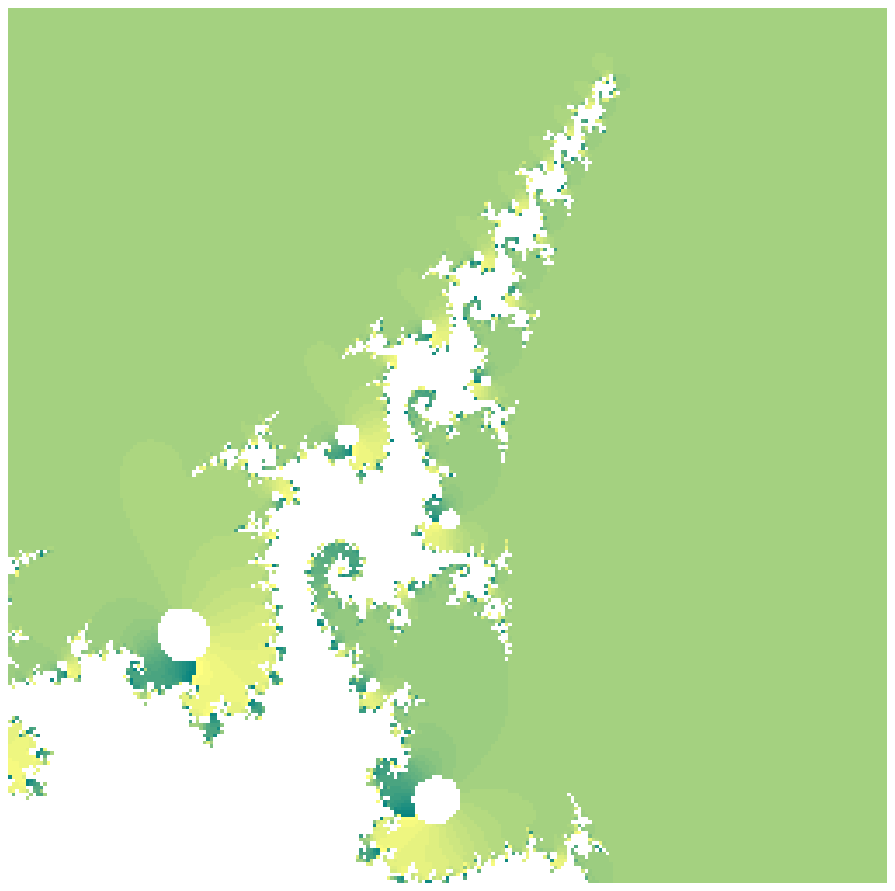}&
  \includegraphics[width=3.8cm]{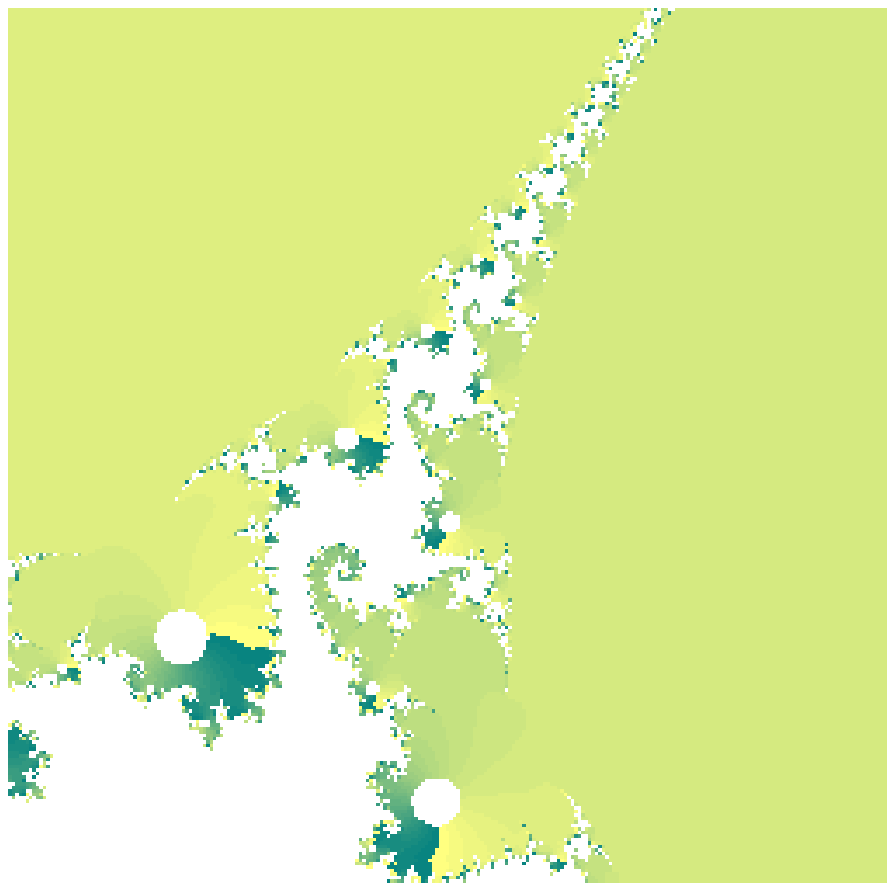}\\
  \footnotesize (A) &\footnotesize (B) &\footnotesize (C) \\
\end{tabular}
\setcaptionwidth{0.9\textwidth}\captionstyle{normal}\caption{\textbf{Wedging
the bounded basin (II)}. Three close-ups of the Julia set $J$ for
the function \eqref{Eq3} with $\theta$ expressed by \eqref{Eq2},
rendered by the holed star model. The iteration index was set at
(A) 150, (B) 300 and (C) 1000 respectively. White disks are
preimages of the hole in the star, while colours are changing due
to iteration values attained by the forward images at different
ranks. The wedging action of the hedgehogs generation is
evident.}\label{Hedge38-40}
\end{figure}

\subsection{Unveiling the code}\label{Unveiling}
Here we explain the elaborations worked by part of the code in
section \ref{HoledBranchedStar}. First one shall focus on this
line:

{\footnotesize
\begin{verbatim}
double out_level = ( f.euclidean_dist( z ) <= offset ) ? -1 : v/potential_rate ;
\end{verbatim}}

\noindent if the point \texttt{z}, resulting from the iteration
process, is farther from the fixed point \texttt{f} than a preset
distance \texttt{offset}, the ratio \texttt{v/potential\_rate}
shall be computed to know whether one paints that iterated point
black or not; otherwise, the negative value $-1$ will be used as a
flag, indicating that the point must not be tested for further
painting. The variable \texttt{offset} stores the radius value for
the hole in the star model.

The value \texttt{v} is the angle of the last image \texttt{z} in
the current orbit and helping to have that point location in relation
to the branches. One also notices that

{\footnotesize
\begin{verbatim} double potential_rate = existence_interval / branches_number ;
\end{verbatim}
}

\noindent which, applied to the first line of code, let \texttt{v / potential\_rate} turn into
$$\footnotesize{\tt index}=\frac{{\footnotesize\tt v\cdot branches\_number}}{{\footnotesize\tt existence\_interval}}=\frac{{\footnotesize\tt v}}{{\footnotesize\tt existence\_interval}}\cdot{\footnotesize\tt branches\_number}$$
the ratio factor in the right member is normalized in the unit interval
[0,1]; when multiplied by
\texttt{branches\_number}, one finds that:
$$\textnormal{\footnotesize 0\ $\footnotesize <$\ {\footnotesize\tt index}\ $\footnotesize <$\ {\footnotesize\tt branches\_number}},$$
so it ranges inside the interval of arbitrary input number of
branches. If decimal values are retrieved, they shall be rounded
off to the integer part, because decimals do not make sense in
terms of branches index. For example, given $31$ branches for an
holed star, like in table \ref{Table05}, we expect that the index
ranges in [0,30]. The product with the \texttt{branches\_number}
yields the branch index for the last computation of the
$n$--iterated image \texttt{z}. It is reasonable that watching
hedgehogs pictures one first wonders: \emph{given a white
background, how come are black lines painted exclusively?} And, in
addition, since the black pixels refer to those only points to be
painted, \emph{why are others not painted?} One side implication
from a general method based upon equi-potentials is that screen
points are plot only if the current equipotential is trespassed.

In our computational model, the geometry locus of the curves is
the qualitative part. This locus is often detected by a formula:
for example $|z-\zeta|=R$ is known to detect a circle centered at
$\zeta$ and if `$=$' replaces with `$<$', we deal with the open
disc.

In order to meet machine requirements operativity, we shall deal
with quality in terms of quantity: therefore, each curve matches
one only index value. When we point out to a curve or to a branch
in the star model, we want its index essentially. With `\emph{an
equipotential level was trespassed}', we mean that the current
index value has \emph{changed}. For a set of points retrieving a
same index, only the first one shall be plot; again, if the
indexes of two consecutive points are different, the second point
is plot. Algorithmically speaking, one stores the last index and,
if it equals the previously stored value, no equipotential is was
trespassed; otherwise, it was and one paints the related point
black. So the implication is straight-forward:
$$\footnotesize\textnormal{\em Different indexes}\rightarrow\textnormal{\em Different equipotential}\rightarrow\textnormal{\em Trespass}\rightarrow\textnormal{\em Paint the point}.$$

For example, if we associate each screen point to its resulting
index, we find a chain of indexes indicating how pixels and the
entire screen row is regarded in this raster process from left to right:
$$\textnormal{\textbf{1}-1-1-\textbf{2}-2-2-2-2-2-\textbf{3}-3-3-3-\textbf{4}-4-4-4-\textbf{2}-2-2-\textbf{5}-5-5-5-\textbf{3}-3-3-3-3}$$
The index in bold \emph{makes the difference} and arrests the
previous sub-sequence of same indexes: the related point will be
painted black. Technically, this can be easily understood by
handling the boolean flag \texttt{bInitFlag} of the main routine
at section \ref{HoledBranchedStar}. The reader would take care
that as the \texttt{bInitFlag} is set, the values for comparison
were stored in the two variables \texttt{val} and
\texttt{tmp\_val}, as shown here in the following resume of the
previous code:

{\footnotesize
\begin{verbatim}
BOOL bDraw = Holed_Star( output_z, fxd_pt, double& val, double&
tmp_val, BOOL& bInitFlag ) ;

if ( bInitFlag )
{
    if ( val != tmp_val )
    {
        if ( bDraw ) pDC->SetPixel( x, y , 0 );

        val = tmp_val ;
    }
}
else bInitFlag = TRUE ;
\end{verbatim}
}

\noindent It is clear that the point is painted black (RGB value
is $0$) only when
\begin{center}
\footnotesize\texttt{if ( val != tmp\_val )}
\end{center}
is satisfied. If so, the old equipotential was trespassed. We
jumped into another equipotential level (iterations are a discrete
model of a continuous map and jumps are solely achieved onto the
successive or onto the preceding equipotential level). And we take
the new value as the referring one by \texttt{val = tmp\_val} ;

\subsection{The pro(s) and con(s)}\label{ProsCons}
\emph{Benefits}: the fate of orbits $f^n(z):z_n$, which finally
land on a same equipotential level $\mathcal{R}$ -- the bounded
region radially distributed around $\delta$, is grouped by the
seed points $z$ under the same color, according to our approach to
`be-already-there-and-work' (p.
\pageref{be-already-there-and-work}). From the early works by
Cherry \cite{Cherry-1964} and the latest production by
P\'{e}rez-Marco \cite {PerezMarco-1997}, hedgehogs dynamics can be
simply resumed as follows (\cite{Cherry-1964}, p. 33):
\begin{quotation}
\small\em ``\dots as it progressively deformed through starfish
shapes \textnormal{[\dots]} consisting of an infinity or `rays'
emanating from $O$; each such `ray' is a connected closed set, and
`most' of them are arbitrarily short. A $z_0$ on any of these rays
gives a chain \textnormal{($z_n$)} each of whose points lies on
another of them \dots''
\end{quotation}

\begin{figure}[h]
\centering
\begin{tabular}{cccc}
  \includegraphics[width=2.8cm]{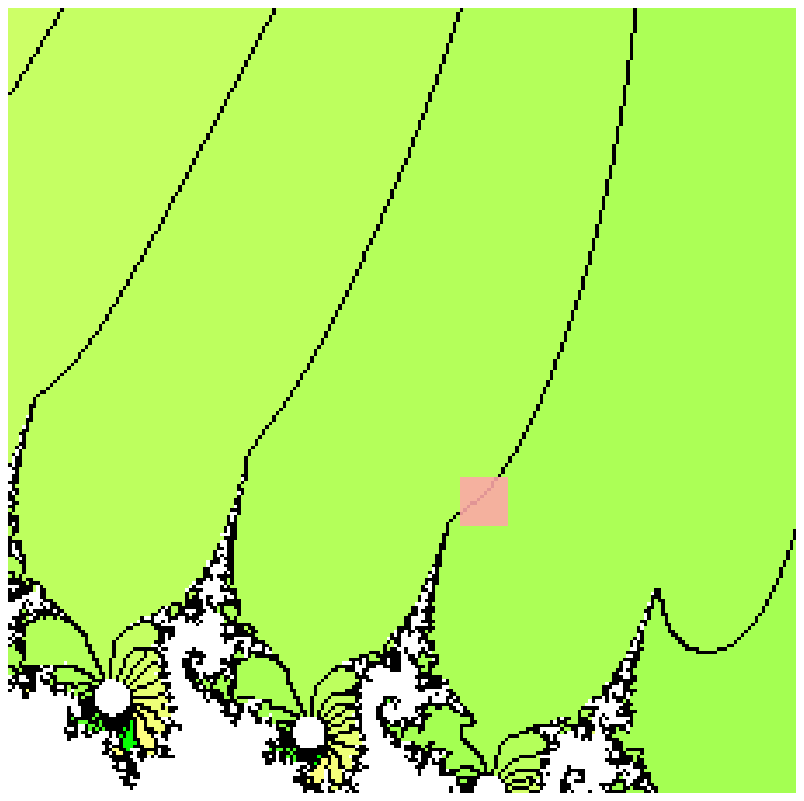}&
  \includegraphics[width=2.8cm]{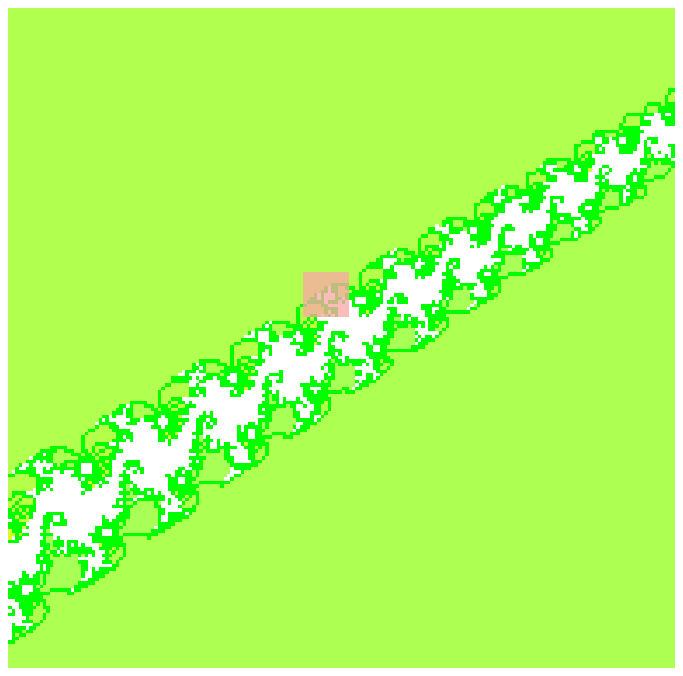}&
  \includegraphics[width=2.8cm]{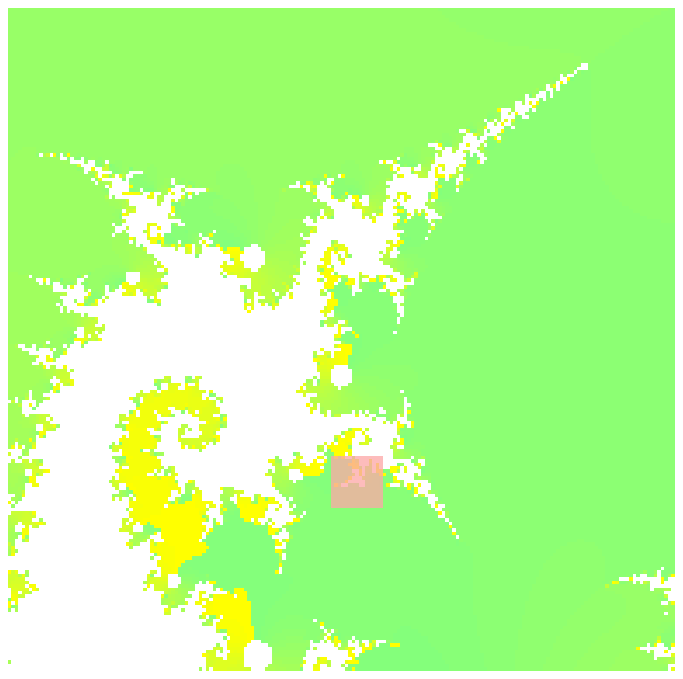}&
  \includegraphics[width=2.8cm]{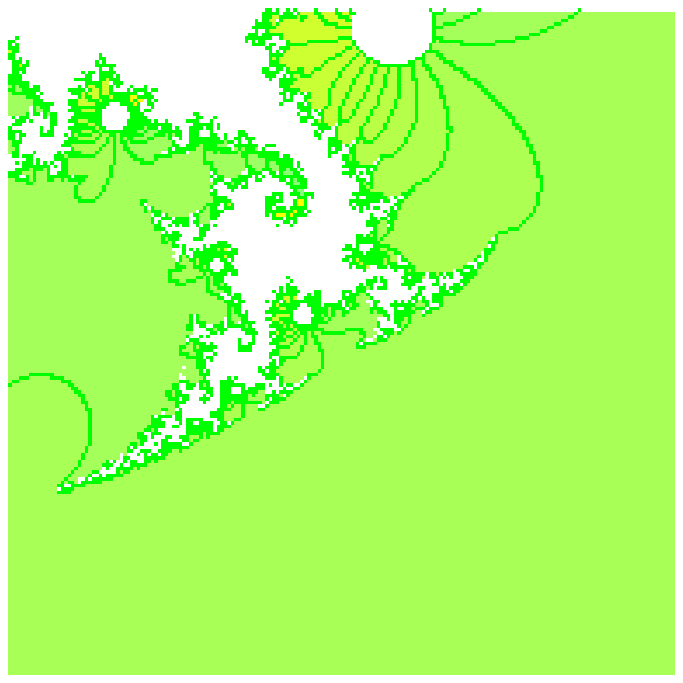}
  \\
  \footnotesize (A) &\footnotesize (B) &\footnotesize (C)&\footnotesize (D)\\
\end{tabular}
\setcaptionwidth{0.9\textwidth}\captionstyle{normal}\caption{\textbf{Wedging
the bounded basin (III)} By zooming in the red square region in
succession, we follow the wedging action of the hedgehog inside
the basin boundary. (A) without requiring a huge number of
iterates (just 100 here) the contour is able to show how the
hedgehog makes its way into the basin. (B--D) There are several
smaller and smaller fjords emanating from the longest arm which
runs towards the linearization domain (whether it is empty or
not).}\label{Hedge41-44}
\end{figure}

Hence the hedgehog boundary $\mathcal{H}$ of the regions
$\mathcal{R}$ wedges the bounded basin, which gets thinner and
thinner as the iterative index $n\rightarrow\infty$. The final
shades distribution helps to evince the topology of $\mathcal{H}$
(figs. \ref{Hedge38-40}), but a major benefit comes from the
contour plot of all such $\mathcal{R}$ \footnote{Which is the same
as plotting the contour of the star itself, in equi-potential
terms.}; contours allow to watch the complete hedgehog topology,
by a comparatively little iterative index, while other methods, even
if pushed to the top machine performance, would have not reached so far.

Good pictures on the way $\mathcal{H}$ wedges the bounded basin are
granted because of the holed star fits the hedgehogs topology, without
needing to input very huge indexes for having almost the same results
which previous standard methods: in fact they suffer the very slow
accumulation rate of neighboring orbits $z_n$ about $\delta$.
This achievement completely fills the lack of the approaches
discussed before.

\emph{Lacks}: more than any other equipotential model, the holed
star does not grant an `intelligent approach', that is, capable of
fitting any input function $f(z)$ as best. Equi-potential curves
are just as a sort of elastic sheet being stretched by $f(z)$:
they are used to watch the local dynamics via lines
distribution, but an a priori knowledge about the property of
given dynamics is required to choose the best model to depict them,
among a bunch featuring a given
equi-potential curves distribution (horizontal or vertical
straight-lines, concentric discs, stars), each one working differently
on the current case, thus being restrictedly indicative on the
function behavior. Hence we may look at equi-potential models
are differently styled interpreters and,
like for any interpretative process, one must
speak the same `language' adopted, here of the dynamics of $f(z)$,
for best understanding: so the essence of any model is to play
as both the proper syntax and as the fittest dictionary to achieve
the correct `translation'. Hedgehogs and holed star model do not
escape this rule, so that the fine tunings can secure the most
adherent results. This is not so difficult however, anyway the star
model is less `ready-to-go' than concentric circles or straight
lines distributions of equi-potentials, because of one more
parameter to tune, namely the hole radius, in order to properly
plot the variety of hedgehogs, as listed at p.
\pageref{HedgehogsList}.

\subsection{Tuning the disc radius}\label{TuningTheRadius}
\begin{figure}
  \centering
\begin{tabular}{ccc}
  \includegraphics[width=4.5cm]{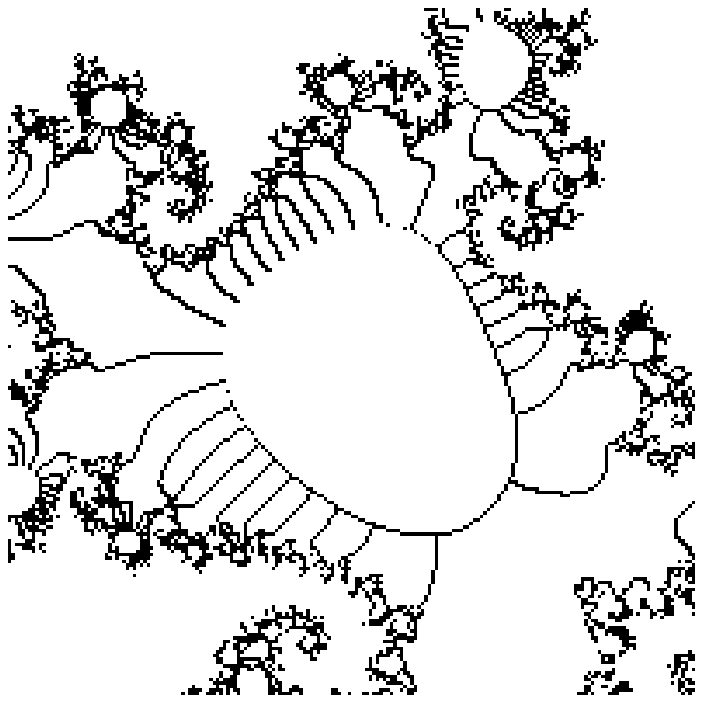} &
  \hspace{0.5cm} &
  \includegraphics[width=4.5cm]{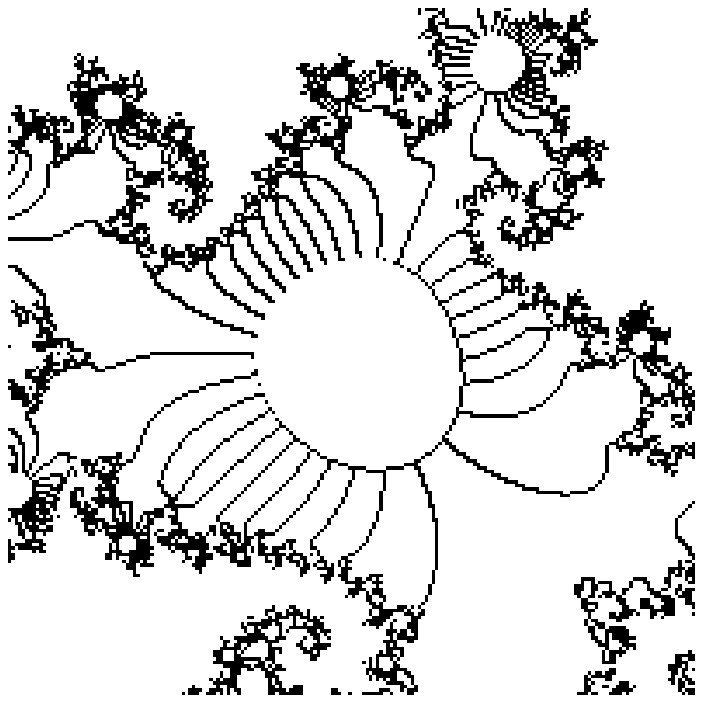} \\
  \footnotesize (A) $r_{\mathcal{S}}=0.20$ & & \footnotesize (B) $r_{\mathcal{S}}=0.15$ \\
  \vspace{0.2cm} & \\
  \includegraphics[width=4.5cm]{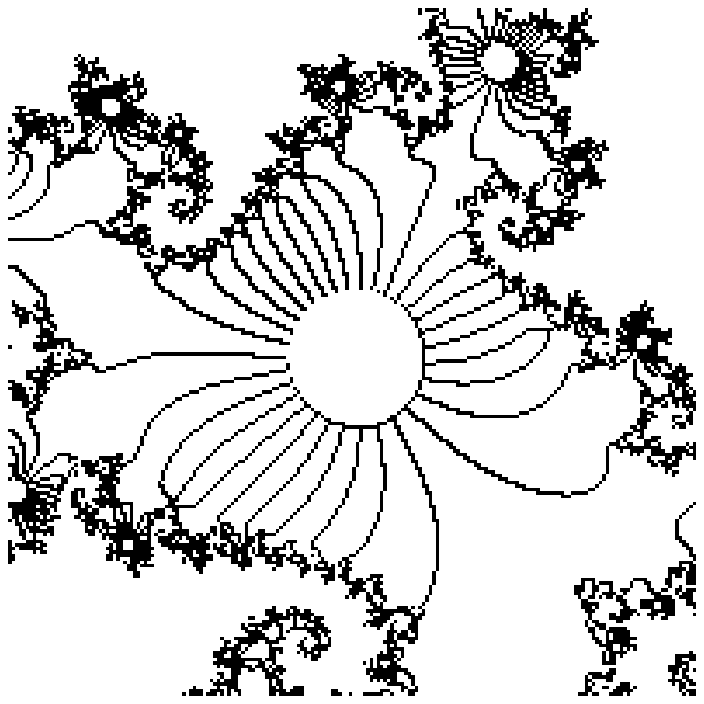} &
  \hspace{0.5cm} &
  \includegraphics[width=4.5cm]{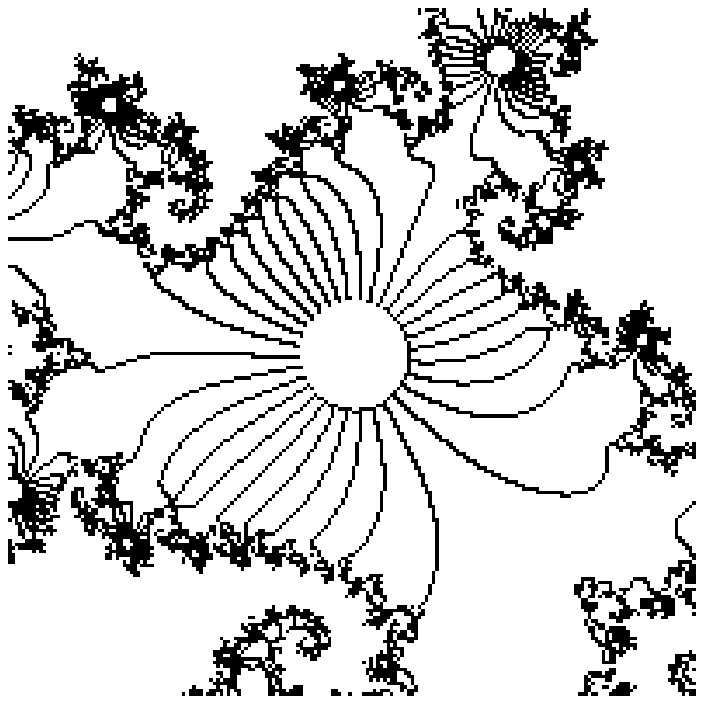} \\
  \footnotesize (C) $r_{\mathcal{S}}=0.10$ & & \footnotesize (D) $r_{\mathcal{S}}=0.09$ \\
\end{tabular}
\caption{\textbf{The radius test}. Tuning the radius
$r_{\mathcal{S}}$ of Siegel compactum.}\label{Table05}
\end{figure}

Being
this one an empiric approach, one needs to tune some parameters to
`grasp' the finest pictures of hedgehogs, letting their characters
to match with the real figure as closer as possible. In figures at
table \ref{Table05}, one tunes the maximal disc/hole radius
$r_{\mathcal{S}}$ for which the hole of our branched star model
would look like a disc: the response from too large values (fig.
\ref{Table05}/A) offers a strongly deformed curve indeed, while
smaller ones may not evince the effective extension of the Siegel
compactum. Thus one has to try to decrease the radius value
$r_{\mathcal{S}}$ until the disc comes out.

The author is working on a theory to
retrieve sharp estimation of such radius metric, as well as for the
Siegel disc \cite{Rosa-Accessibility}: this might be helpful to
finally have very fine pictures of hedgehogs.

\subsection{Stars and equivalent classes}\label{StarsClasses}
As shown, orbits accumulate at their forward invariant limit sets.
In the case of hedgehogs, whether the linearization disc is small
or has zero area, the holed $k$-th branched star is a model
describing this accumulation in the same way as neighboring orbits
do. The star model benefits are two-fold: while the regions
imitate as best as possible such dynamics inside the bounded
basin, the branch lines are screening the shape of the basin
$\mathcal{B}_\infty$ to $\infty$. Since the latter and the bounded
basin are complementary sets, in terms of converging orbits,
tracking down their shapes equals to know the hedgehog topology.

With regard to the star and the regions between, orbits are parted
into equivalence classes
$$\mathcal{E}_1,\mathcal{E}_2,\dots,\mathcal{E}_k,$$
where $k$ is the number of the branches. The basin
$\mathcal{B}_\infty$ to $\infty$ does not belong to any class
$\mathcal{E}$, thus it does not belong to
$\displaystyle\bigcup_{n=1}^k\mathcal{E}_n,$ since the union is
the bounded basin $\mathcal{B}_\delta$ itself. In the holed star
model, $\mathcal{B}_\delta$ is mapped to the union of the regions
between the branches.

The branches of the star model elongate from the outer region up
to the disc, or to the center point if the hole has zero area:
this is not merely obvious and plays relevantly during orbits
classification. In fact, in a similar fashion but applied to
iterates, the imitation of the branches distribution allows to
track down the wedging action of $\mathcal{B}_\infty$ up the
boundary of the hole or, again, to the center of the star when the
hole is empty: even here, consequentially, the wedging action by
$\mathcal{B}_\infty$ to $\mathcal{B}_\delta$ helps to deduce the
hedgehog shape (\emph{This is the reason why one should already
know the possible existence of the hole, i.e. of a linearization
area with positive radius.}) As remarked differently in section
\ref{ProsCons}, one great benefit is the imitation \emph{up to the
hole boundary}. In fact, nearly iterates get slower and slower to
converge or even their might not converge at all for hedgehogs.
The imitation of branches distribution allows to break down the
slow convergence barrier and the resulting line show the orbits
fate, even after a relatively\footnote{But the smaller
neighborhoods to blow up are, the finer they need to be displayed
and thus the larger iteration index shall be.} slow number of
iterates (see figs. \ref{Hedge41-44}).

In fact the equivalence classes allow to reconstruct the basins
distribution in the same fashion of the star model: so one also
understands why it also works finely with the Fatou-Leau flower, a
similarly shaped vector field, where one basin wedges the other.

\subsection{Displaying the Fatou-Leau flowers}\label{DisplayingFatouLeau}
\begin{figure}
        \centering
        \begin{tabular}{ccc}
            \includegraphics[width=5.6cm]{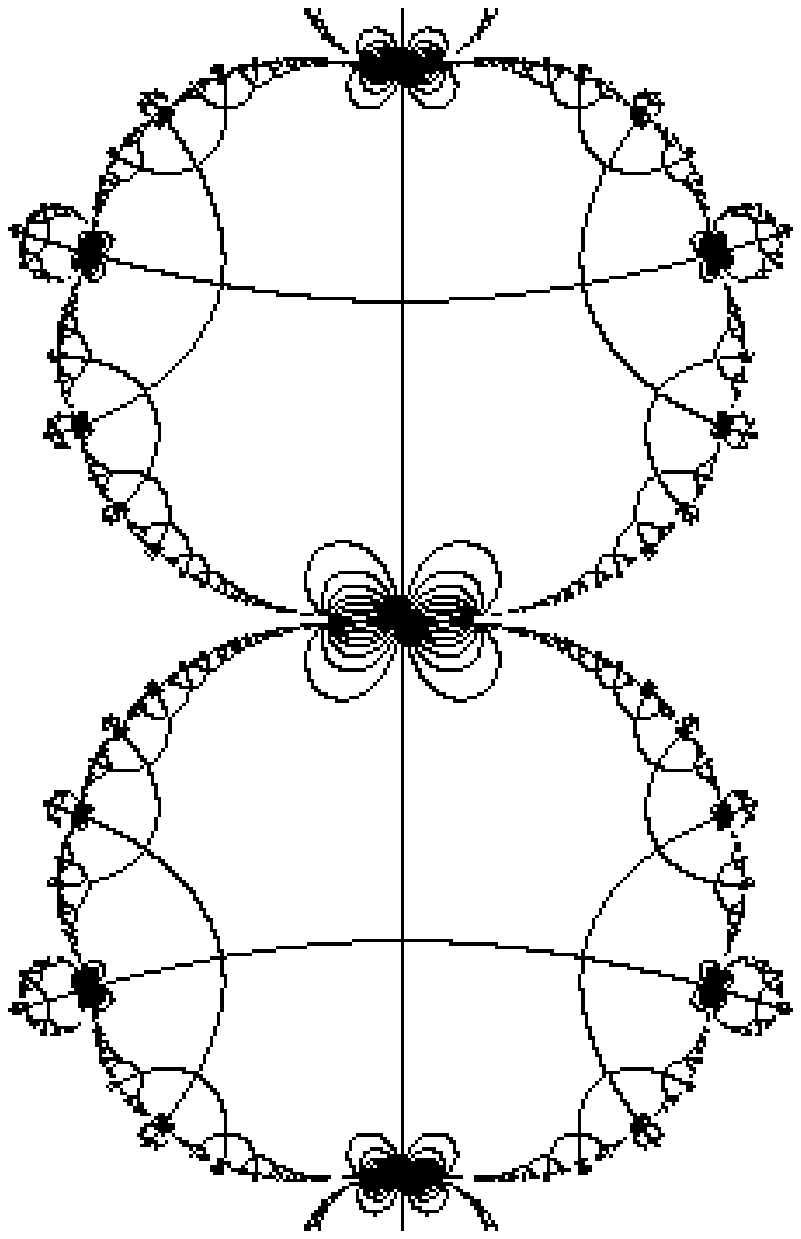} & &
            \includegraphics[width=5.6cm]{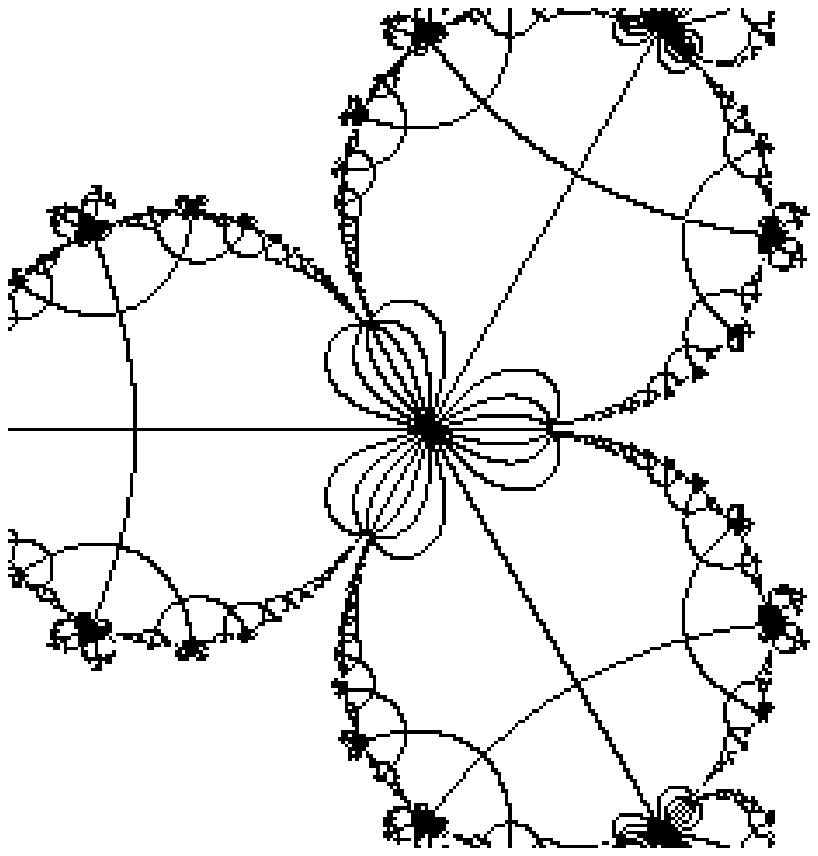}\\
            $z+z^3$ & & $z+z^4$\\
            \vspace{0.3cm}\\
            \includegraphics[width=5.6cm]{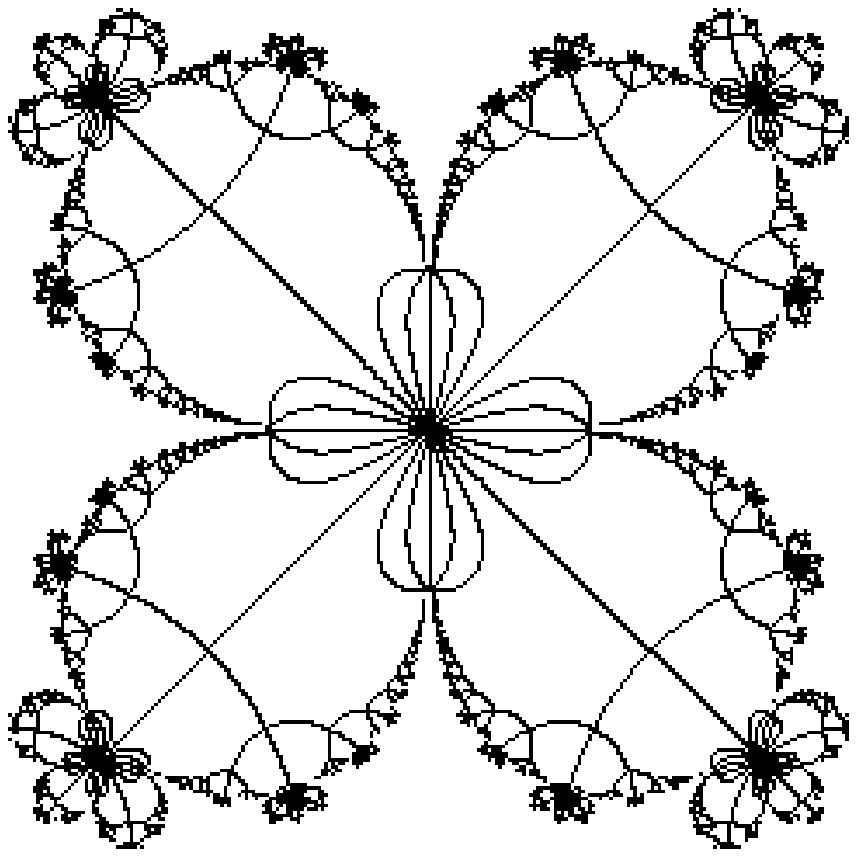} & &
            \includegraphics[width=5.6cm]{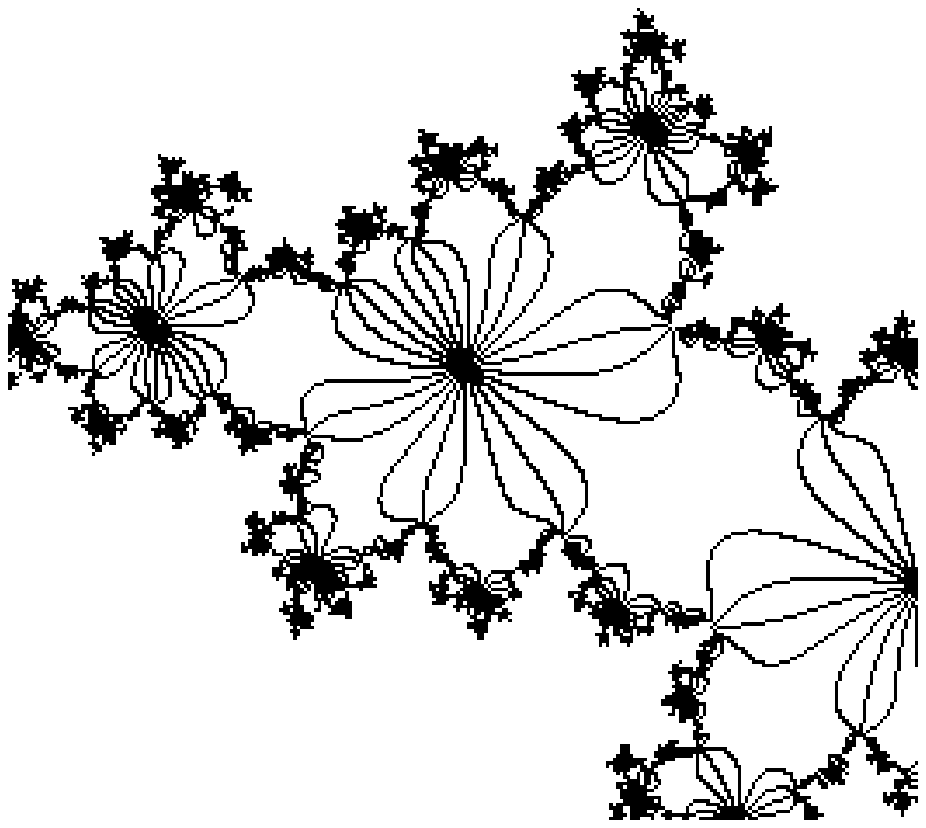}\\
            $z+z^5$ & & $e^{2\pi it}z+z^2,\quad t=3/7$\\
        \end{tabular}
\setcaptionwidth{0.9\textwidth}
\captionstyle{normal}\caption{\textbf{A bunch of flowers to the
star}. The application of the holed star to different cases of
Fatou-Leau's flowers.}\label{SideTable01}
\end{figure}

The holed star immediately worked very finely with topologically
equivalent invariant sets, so one might like to consider the
benefits from this new graphical construction and focus on the
Fatou-Leau flowers in table \ref{SideTable01} before moving definitely to
hedgehogs.
\begin{definition}\label{DefFatouLeauFlower}
\textbf{Fatou-Leau flower.}\ If $\hat{z}$ is a fixed point of
multiplicity $n+1\geq 2$, then there exist attracting petals
$\mathcal{P}_1, \mathcal{P}_2, \dots, \mathcal{P}_n$ for the $n$
attracting directions at $\hat{z}$, a repelling petals
$\mathcal{P}'_1, \mathcal{P}'_2, \dots, \mathcal{P}'_n$ for the
$n$ repelling directions, so that the union of these $2n$ petals,
together with $\hat{z}$ itself, forms a neighborhood $N_0$ of
$\hat{z}$. \textnormal{(\cite{Milnor}, p. 105)}
\end{definition}

One first notices that straight lines meet at the origin and they
lie at the attracting directions, whereas the divergent ones are
deformed into petals shaped curves.

It is useful to remark that straight lines leading to the origin
are adjacent to converging orbits: the imposition of the holed
star model lessens the difficulty of reaching closer neighborhoods
of $\delta$, owing to the slow convergence rate inside. Thus the
computation and display of such flowered invariant set
may be a good entry point for realizing the benefits of such
a qualitative approach and the need of tuning properly the parameters
of the star, such as the radius hole (set to $0$ for flowers) and the
branches number, especially for hedgehogs.

\begin{sidewaystable}
  \centering
  \begin{tabular}{ccccc}
    \includegraphics[width=4.0cm]{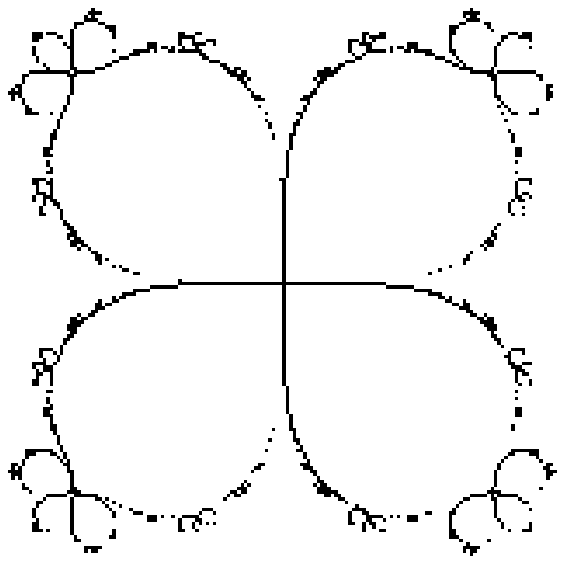} & &
    \includegraphics[width=4.0cm]{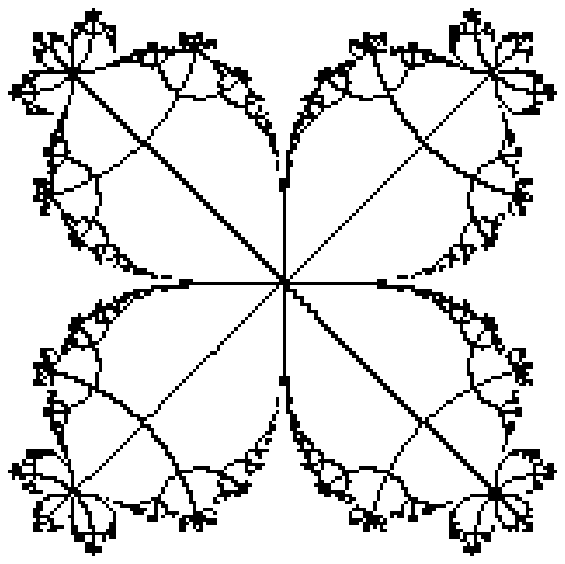} & &
    \includegraphics[width=4.0cm]{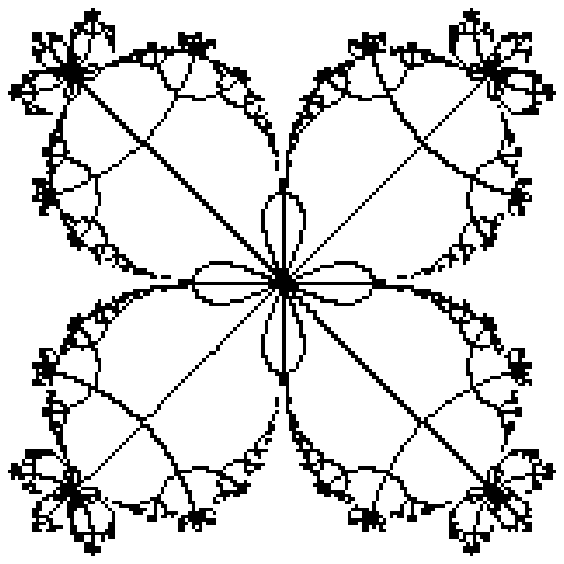} \\
    (A) & & (B) & & (C)\\
    \vspace{0.3cm}\\
    \includegraphics[width=4.0cm]{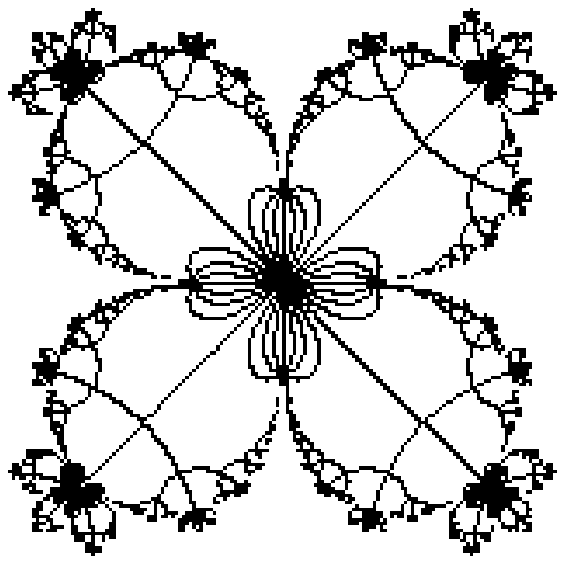} & &
    \includegraphics[width=4.0cm]{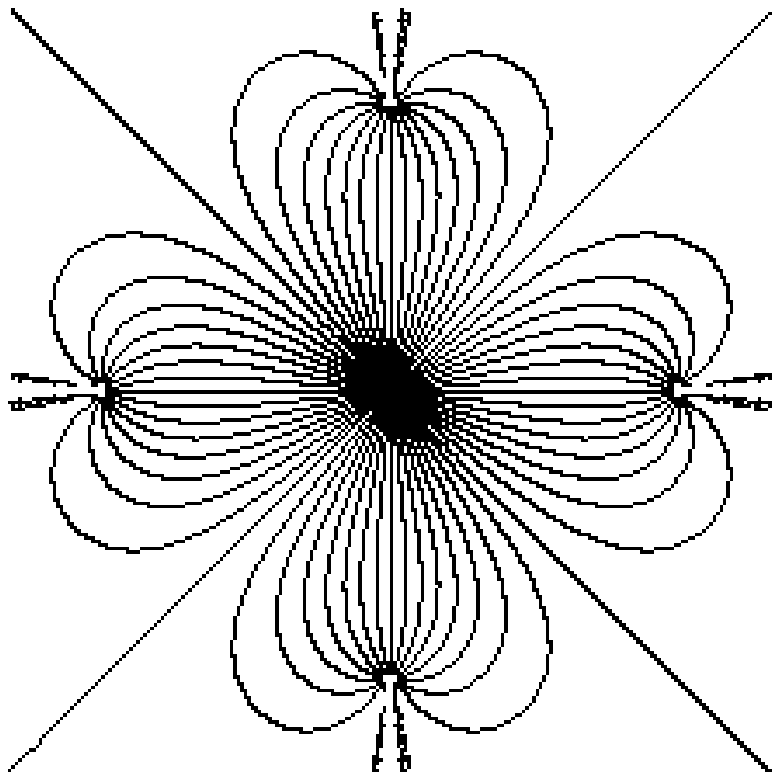} \\
    (D) & & (E)\\
  \end{tabular}
\setcaptionwidth{0.9\textwidth}
\captionstyle{normal}\caption{\textbf{Tuning the number of
branches}. We chose for this example the function $f(z):z+z^5$,
  which yields a flower with $4$ petals. The best performance for the holed branched star
  is to have some branches adjacent to both the $4$ converging and the $4$ attracting directions;
  then it is wise to set the branches number to multiples of $4$. The other branches are useful to understand the dynamics. Without them we should just see straight
  lines, as depicted in (A). So we have 4, 8, 16, 32 and 64 branches for figures (A), (B), (C), (D), (E) respectively.}\label{SideTable02}
\end{sidewaystable}

\subsection{Experiments with hedgehogs}\label{ExperimentsHedgehogs}
This latter aspect should be greatly taken into account for
hedgehogs too.

\cite{Milnor} (p. 123) was rather inspiring to try enhancing their
graphics by focusing on an imitation graphical model based upon
equi-potentials; this book also includes one formula and one
figure of a quadratic holomorphic germ providing an hedgehog. Here
the irrational angle $\theta$ can be expressed in terms of the
continuous fraction expansion
\begin{equation}\label{Eq2}
\frac{1}{3+\frac{1}{10+\frac{1}{20000+\dots}}}.
\end{equation}

\begin{figure}
  \centering
\begin{tabular}{ccc}
  \includegraphics[width=3.8cm]{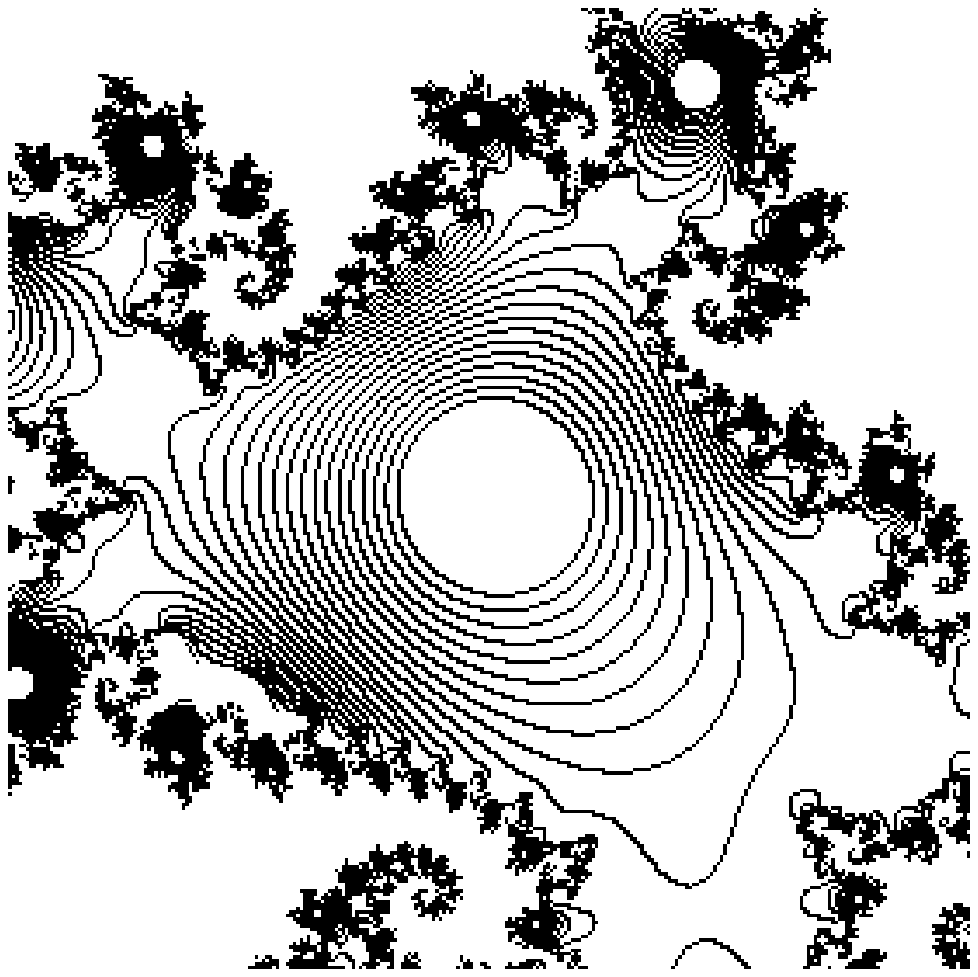} &
  \includegraphics[width=3.8cm]{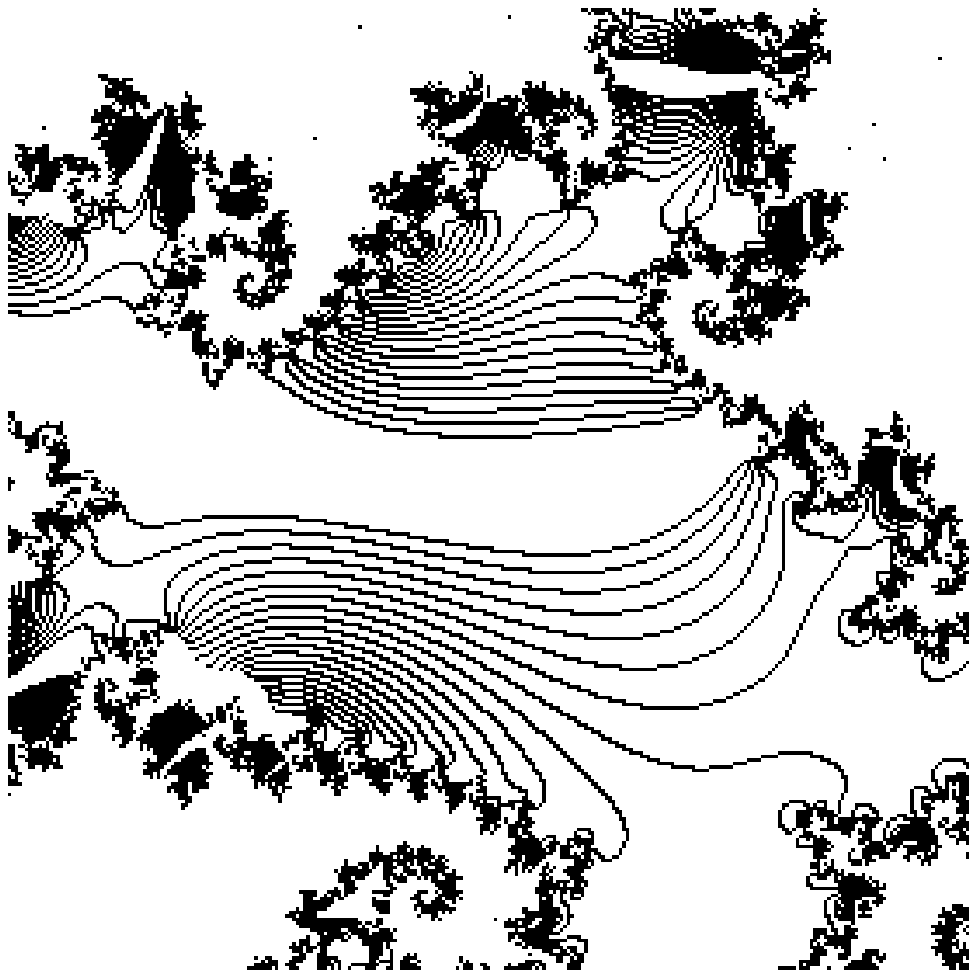} &
  \includegraphics[width=3.8cm]{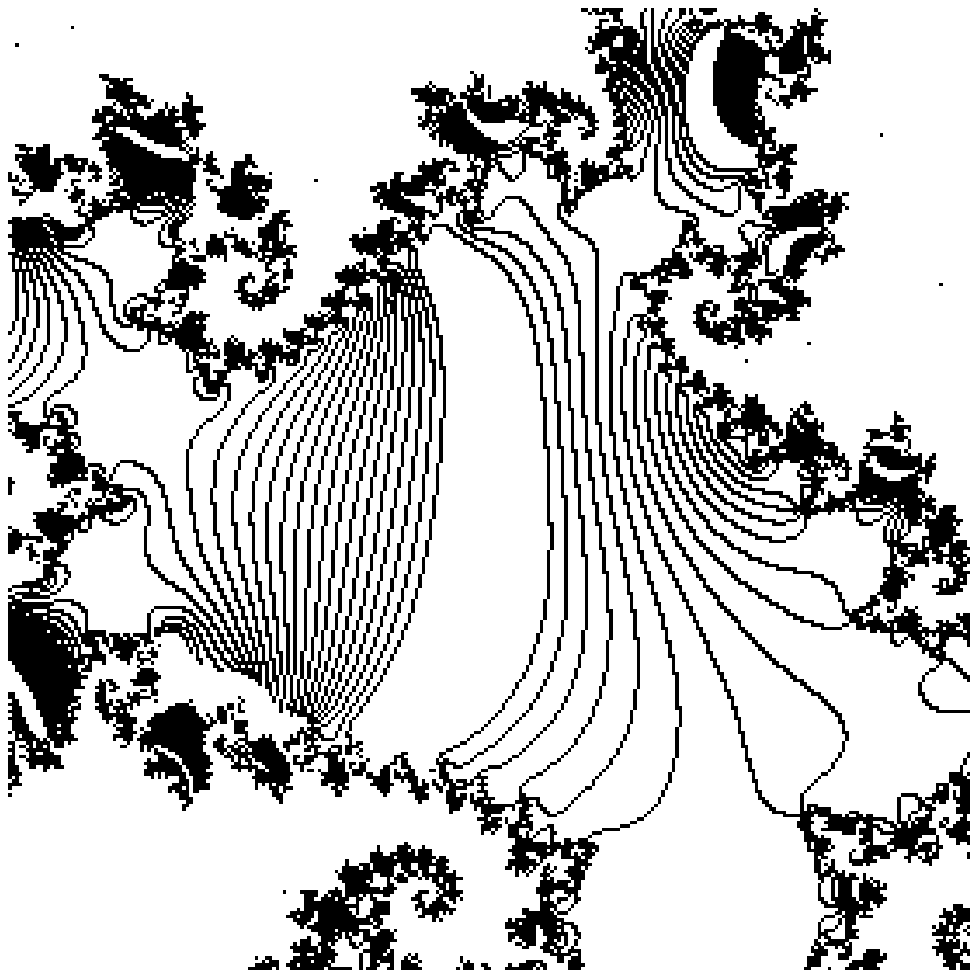}\\
  \footnotesize (A) & \footnotesize (B) & \footnotesize (C)\\
\end{tabular}
\setcaptionwidth{0.9\textwidth}
\captionstyle{normal}\caption{\textbf{Different equi-potential
models}. The failures to evince the hedgehog shape: concentric
disks, vertical and horizontal lines distributions are displayed
in figures (A), (B) and (C) respectively. Finest figures need the
most equivalent equi-potential model.}\label{Table09}
\end{figure}

The choice of such model was not at random: the different
approaches in table \ref{Table09} lack of displaying the true
shape, so that at last the holed star was chosen to best
\emph{imitate} the general hedgehog shape under iteration, when
the Siegel compactum is empty or even arbitrarily small. At this
concern, setting the right number of branches improves the figure.
Our approach wants to exploit the equi-potentials method by
looking at the deformation of the (holed) $n^{th}$-branched star.
One notices the existence of
curves indicating the wedging action into
wedging $\mathcal{B}_\delta$; this was not evident from
figures \ref{Table03}/C and D by ordinary methods. Although we
cannot depict a detailed shape of fjords, we may at least
understand where they will get to.

As we showed, this is not an intelligent method. Thus finest drawings
require to already know the hedgehogs features, like the the radius
of the Siegel compactum and number of fjords, regions where
$\mathcal{B}_\infty$ penetrates $\mathcal{B}_\delta$) and modelled
by the star branches.

\begin{figure}
  \centering
  \begin{tabular}{ccc}
    \includegraphics[width=3.8cm]{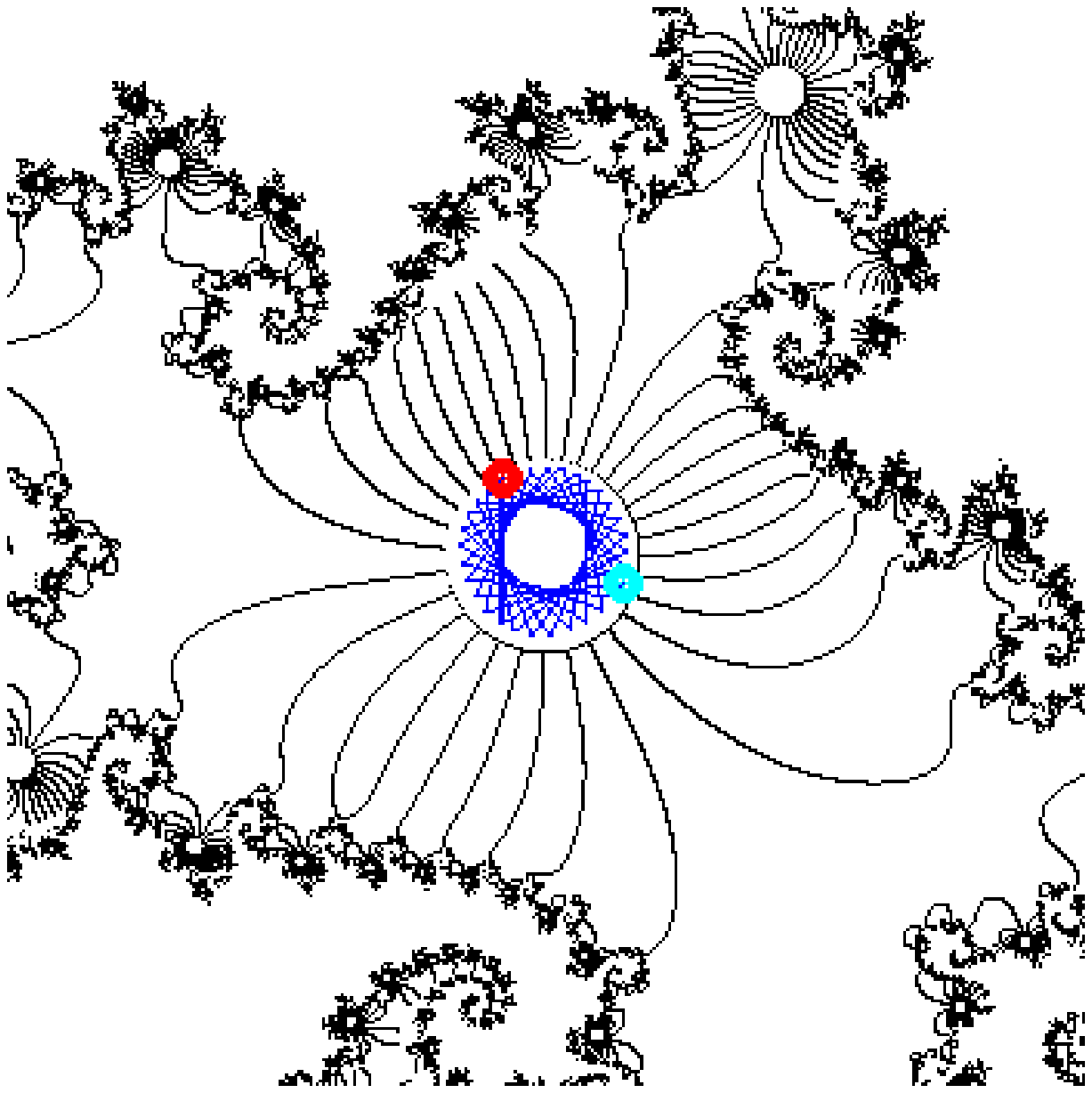} &
    \includegraphics[width=3.8cm]{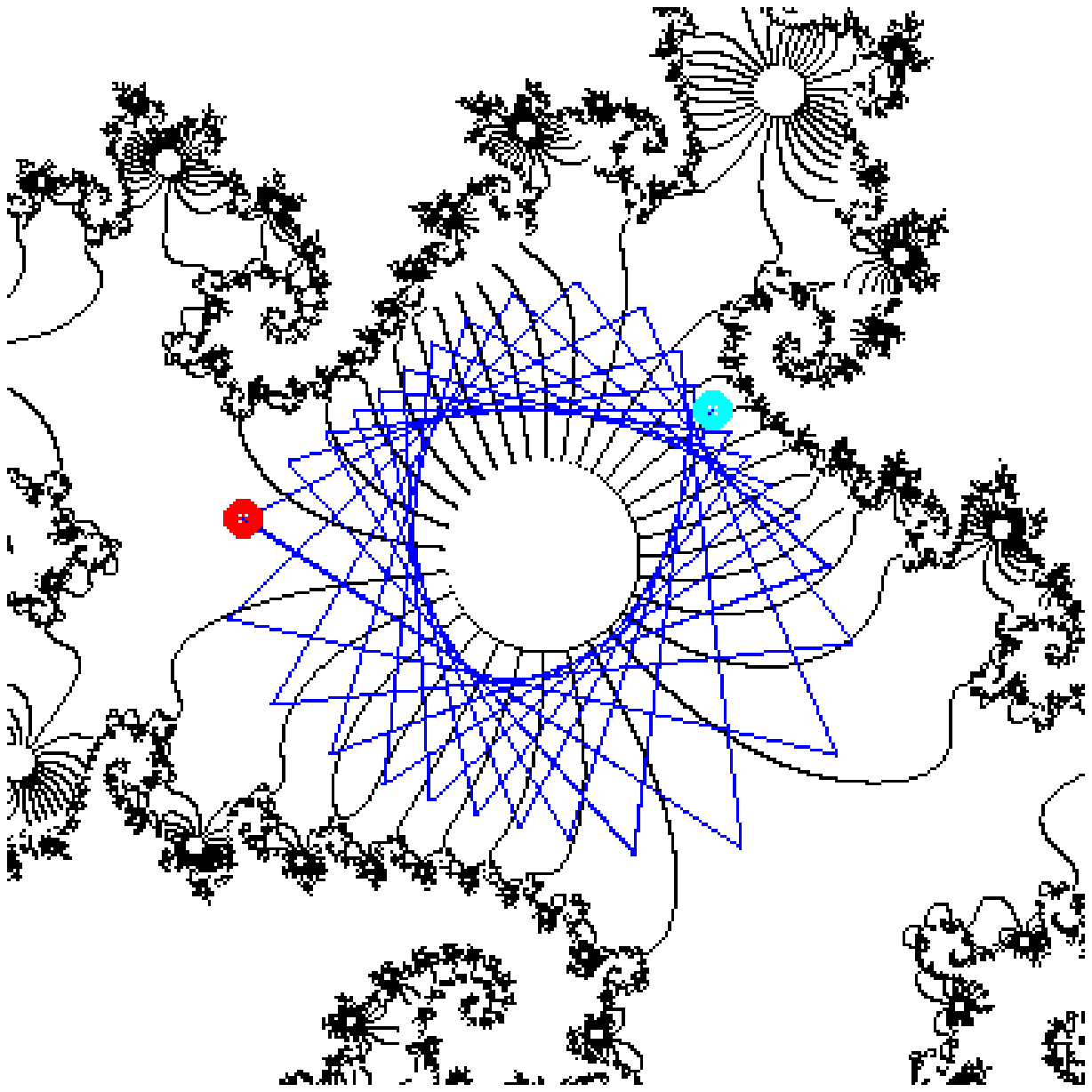} &
    \includegraphics[width=3.8cm]{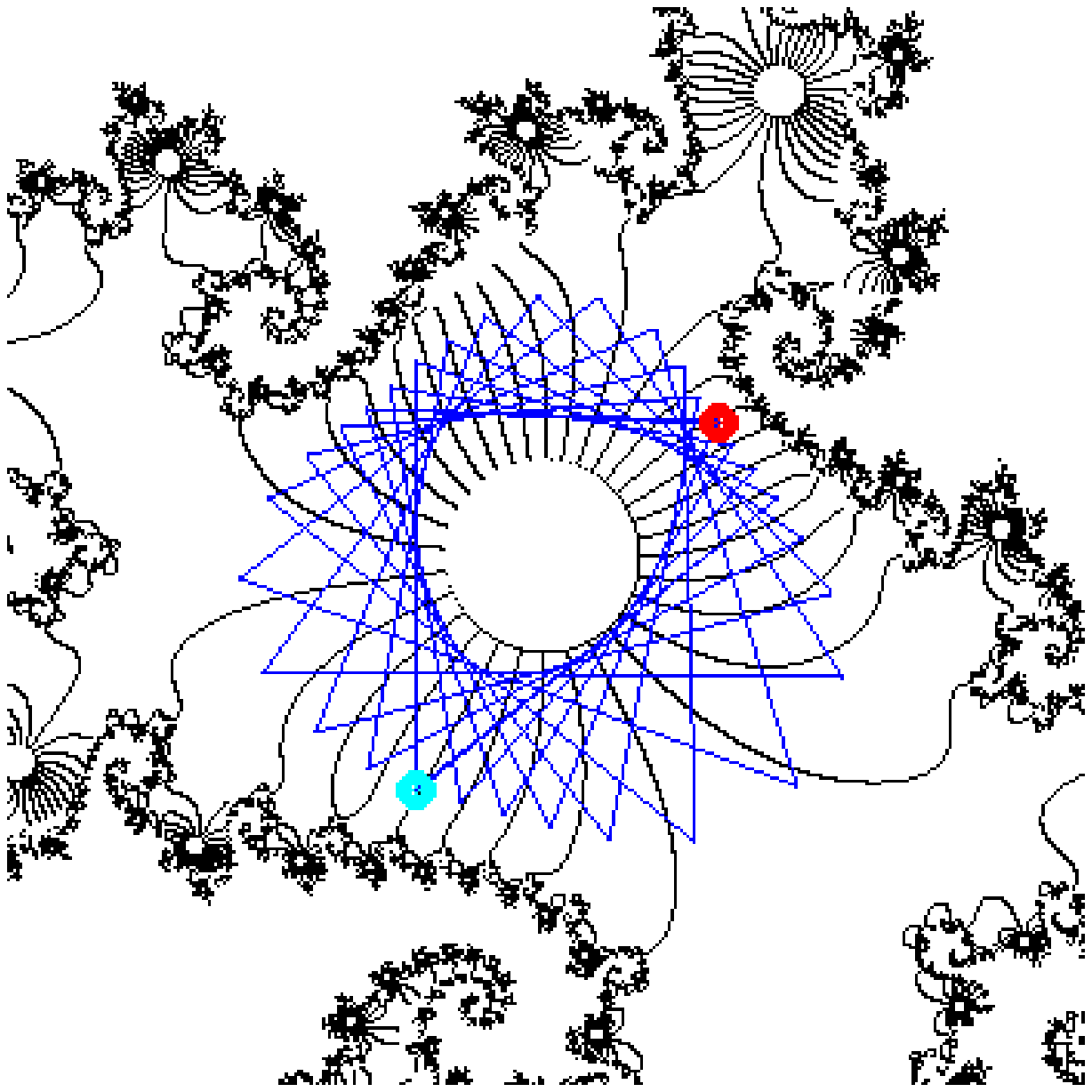}\\
    (A) & (B) & (C)\\
  \end{tabular}
\setcaptionwidth{0.9\textwidth}
\captionstyle{normal}\caption{\textbf{Neighboring dynamics}.
Figure (A) shows the rotatory motion of iterates in the Siegel
compactum, while outer orbits are shown in (B) and
(C).}\label{SideTable03}
\end{figure}

\section{Conclusions}\label{Conclusions}
In author's opinion, well-drawn figures of local invariant sets
necessarily want customized approaches: equi-potential approach
is looking like the best performing today and is could be even
optimized when topologically equivalent model are applied;
thus a general method to be applied for displaying
a given local invariant set, independent on the input map,
seems a utopia. But, being open to the possibilities offered
by Science, one optimistically says `\emph{up to now !}'

\begin{flushright}
    \begin{tabular}{p{4.0cm}}
    \small Alessandro Rosa\\
    \footnotesize\texttt{zandor\_zz@yahoo.it}
    \end{tabular}
\end{flushright}

\footnotetext{Author Rosa's site : http://www.malilla.supereva.it}

\bibliographystyle{amsplain}

\end{document}